\documentclass[review]{elsarticle}

\usepackage{lineno,hyperref}
\usepackage{tikz-cd}
\usepackage{dsfont}
\usepackage{amsfonts}
\usepackage{amsmath}
\usepackage{slashed}
\usepackage{graphics}
\usepackage{amsmath}
\usepackage{amssymb}
\usepackage{tikz-cd}
\usepackage[compat=1.1.0]{tikz-feynman}
\usepackage{setspace}
\usepackage{float}
\usepackage{cleveref}
\usepackage{units}
\modulolinenumbers[5]

\journal{Galaxies (MDPI)}
\begin{document}

\begin{frontmatter}

\title{\textbf{Quantum gravity phenomenology induced in the propagation of UHECR, a kinematical solution in Finsler and generalized Finsler  spacetime}}

\author[Unina,Unimi]{M.D.C. Torri\corref{mycorrespondingauthor}}
\cortext[mycorrespondingauthor]{Corresponding author}
\ead{marco.torri@unimi.it\\ marco.torri@mi.infn.it}

\address[Unina]{Dipartimento di Fisica, Universit\'a degli Studi di Napoli}
\address[Unimi]{Dipartimento di Fisica, Universit\'a degli Studi di Milano and INFN Milano}

\begin{abstract}
It is well-known that the Universe is opaque to the propagation of Ultra-High-Energy Cosmic Rays (UHECRs) since these particles dissipate energy during their propagation interacting with the Cosmic Microwave Background (CMB) mainly in the so-called GZK cut-off phenomenon. Some experimental evidence seems to hint at the possibility of a dilation of the GZK predicted opacity sphere. It is well-known that kinematical perturbations caused by supposed quantum gravity (QG) effects can modify the foreseen GZK opacity horizon. The introduction of Lorentz Invariance Violation (LIV) can indeed reduce, in some cases making negligible, the CMB-UHECRs interaction probability. In this work we explore the effects induced by modified kinematics in the UHECRs phenomenology from the QG perspective. We explore the possibility of a geometrical description of the massive fermions interaction with the supposed quantum structure of spacetime in order to introduce a Lorentz covariance modification. The kinematics is amended modifying the Dispersion Relations (DRs) of free particles in the context of a covariance-preserving framework. This spacetime description requires a more general geometry than the usual Riemannian one, indicating for instance the Finsler construction and the related generalized Finsler spacetime as ideal candidates. Finally we investigate the correlation between the magnitude of Lorentz covariance modification and the attenuation length of the photopion production process related to the GZK cut-off, demonstrating that the predicted opacity horizon can be dilated even in the context of a theory that does not require any privileged reference frame.\\
\end{abstract}

\begin{keyword}
\emph{Quantum Gravity, Lorentz Invariance Violation, Doubly Special Relativity, Finsler Geometry, Astroparticle Physics, Cosmic Rays}
\end{keyword}

\end{frontmatter}

\setcounter{section}{0} 

\section{Introduction}

The Universe is opaque to the propagation of radiation in particular of its most energetic component, that is Cosmic Rays (CR). This part of the extraterrestrial radiation is composed of charged particles, protons and bare nuclei, most likely of extragalactic origin, with energies above $E\sim \unit[1]{EeV}$. Since the Universe opacity is more pronounced at the highest energies, in this work we deal with Ultra High Energy Cosmic Rays (UHECRs) that is the most energetic fraction of CR with energies larger than $\sim \unit[5\times10^{19}]{eV}$. UHECRs during their propagation may interact with the Cosmic Microwave Background (CMB) and may be attenuated in a way that depends on their energy and nature. Protons dissipate energy via pair production or photopion creation, instead bare nuclei can undergo photodissociation processes. This means that the UHECRs free propagation path is finite and if they propagate for long enough distances they can be detected on Earth only under a determined energy threshold. This phenomenon poses an upper limit on the detectable UHECRs energy and is named GZK cut-off from the name of the physicists Greisen, Zatsepin and Kuzmin \cite{Greisen,Zatsepin}.\\
Astroparticle physics can be a useful framework to conduct searches for possible departures from the Lorentz covariance since in some theoretical models the Quantum Gravity (QG) effects are supposed to be more visible in the high-energy limit as Lorentz Invariance Violation (LIV) effects. Currently there is no a definitive theory that unifies quantum physics with gravity. The greatest challenge in formulating such a unified theory is the actual impossibility of obtaining the energies needed to probe the Planck scale realm. It is commonly believed that the Planck energy $E_{P}=\sqrt{\hslash c^{5}G^{-1}}\simeq \unit[1.22\times10^{19}]{GeV}$ represents the threshold that separates the classical formulation of physics, which is General Relativity (GR), from the quantum realm, represented by Quantum Field Theory (QFT). Nevertheless some QG signatures may become manifest in a low-energy regime as residual effects and these perturbations of standard physics may give access to the possible phenomenology induced by QG.\\
The physics correlated with the UHECR propagation can indeed be exploited to investigate new phenomena, such as the supposed quantum structure of the spacetime \cite{Coleman, Scully, Stecker, TorriUHECR,Torri:2020fao,TorriPhD}. UHECRs propagate for cosmic distances and reach extremely high energies, therefore they can open a window on the Planck scale physics.\\
In this work we investigate the possibility to detect the supposed QG signatures as departures from the Lorentz covariance in the UHECRs sector in the context of a theoretical framework that does not require the introduction of a privileged reference frame. There are many theoretical models that investigate the possibility of modifications as QG-induced perturbations \cite{Koste1,Kostelecky2008,Coleman,Cohen,AmelinoCamelia,AmelinoCamelia2,AmelinoCamelia3,AmelinoCamelia4}, but we conduct our investigation in the context of Homogeneously Modified Special Relativity (HMSR) \cite{TorriHMSR}. In this theoretical framework the covariance is modified and not broken as in Doubly Special Relativity (DSR) \cite{AmelinoCamelia,AmelinoCamelia2,AmelinoCamelia3,AmelinoCamelia4}, hence the kinematical symmetry is again valid in an amended formulation. In this context a minimal extension of the Standard Model (SM) of particle physics can be formulated preserving the modified covariance. The QG perturbations are introduced in the free particle kinematics modifying the dispersion relations. This modification is motivated by the idea that the effects induced by the quantum structure of the background can be geometrized requiring a more general structure to set the theory than the usual Riemann geometry, the Finsler one.\\
The phenomenological analysis is limited to the lightest component of the UHECRs, that is protons. This component is interesting since it is the least deflected one by magnetic fields during propagation so proton trajectories are more straightforward than those of the heavier CRs. The introduction of QG kinematical perturbations can modify the allowed phase space for the interaction with the CMB and the pion creation process. The phenomenological effects are evaluated via numerical simulations conducted using an ad hoc version of the software \emph{SimProp} \cite{Aloisio}, modified in order to include the QG kinematical perturbations \cite{Torri:2020fao}. The introduction of Lorentz covariance modifications induce an opacity horizon enlargement \cite{TorriUHECR,Torri:2020fao}, as foreseen even in the case of pure LIV \cite{Coleman, Scully, Stecker}.\\
This work is structured as follows: first we introduce the theoretical framework motivated by the QG perspective, geometrizing the interaction of a free particle with the background. Then we introduce a minimal extension of the Standard Model of particle physics in an amended covariant formulation. In the following we introduce the physics related with the UHECRs propagation explaining where the QG effects can manifest themselves. Afterwards, we present the numerical results obtained via the modified \emph{SimProp} software. Finally, we analyze the results with particular emphasis on the possibility of obtaining detectable phenomenological effects in this sector in the context of a covariant framework, that retains a modified version of the kinematical symmetry group (the Lorentz/Poincaré one).

\section{Kinematical modifications in an isotropy preserving scenario}
In this work we explore the possibility of investigating the presumed QG effects perturbing the free particle kinematics and we obtain a minimal extension of the particle SM in a framework that preserves an amended covariance formulation.

\subsection{Modified Dispersion Relations}
The QG-caused perturbations are introduced only in the free particle kinematics in order to prevent the introduction of any exotic particle or reaction. In this work as a reasonable physical hypothesis only the dispersion relations of massive fermions are modified, because of the gravitational nature of these supposed effects. The QG perturbations are introduced modifying the Dispersion Relations (DRs) of massive particles, as done in the great part of theories \cite{Koste1,Coleman,Cohen,AmelinoCamelia,AmelinoCamelia2,AmelinoCamelia3,AmelinoCamelia4,Smolin1,Smolin2} that
tackle the Special Relativity (SR) modifications from a QG perspective. In this work every particle species is supposed to have its personal Modified Dispersion Relation (MDR):
\begin{equation}
\label{a1}
F^{2}(p)=E^2\left(1+h\left(\frac{\left|\vec{p}\right|}{E}\right)\right)-\left|\vec{p}\right|^2\left(1+k\left(\frac{\left|\vec{p}\right|}{E}\right)\right)=m^2\,,
\end{equation}
with peculiar functions $h$ and $k$ depending on the particle species. These functions have to satisfy the perturbation condition:
\begin{equation}
\label{a1a}
|h|\ll1\qquad |k|\ll1\qquad h,\,k\simeq O(1)\,,
\end{equation}
with derivatives of first and second order limited with at least the same magnitude. One may consider, for instance, the linear function $\alpha(|\vec{p}|/E)$ or the translated exponential $\exp(\alpha|\vec{p}|/E)-1$ with $\alpha$ a strongly limited coefficient that encodes the perturbation character of the selected function. In the following all the terms with comparable magnitude will be considered first-order perturbations\footnote{See appendix A} $\simeq O(1)$.\\
The perturbation functions $h$ and $k$ are 0-degree homogeneous and depend on the 3-momentum magnitude $|\vec{p}|$ in order to be rotationally invariant, as conjectured in \cite{TorriHMSR}. The main motivation underlying this perturbation choice is to guarantee the MDR geometrical origin, in this way the function $F^{2}(p)$ is 2-homogeneous and can be a candidate Finsler pseudo-norm. It is important to emphasize that we are dealing with a pseudo-Finsler geometry since the metric signature is not positive definite, in the sense that the underlying structure is the Minkowski geometry. As it will be clear in the following, if the perturbation functions are chosen as $h=k$ one can obtain a conformally flat momentum space and if $h>k$ the MDR \cref{a1} can be approximated with the MDR defined in \cite{TorriHMSR}:
\begin{equation}
\label{a1b}
F^{2}(p)=E^2-\left|\vec{p}\right|^2\left(1-f\left(\frac{\left|\vec{p}\right|}{E}\right)\right)=m^2\,,
\end{equation}
with $f\simeq(h-k)$. Another important feature emerges thanks to the dependence of the perturbations on the ratio $|\vec{p}|/E$, indeed with the latter MDR choice:
\begin{equation}
\label{a2}
\begin{split}
&\frac{|\vec{p}|}{E}\rightarrow(1+\delta)\;\;\text{for}\;\;E\rightarrow\infty\;\;\text{with}\;\;0<\delta\ll1\\
&f\left(\frac{|\vec{p}|}{E}\right)\rightarrow\epsilon\ll1
\end{split}
\end{equation}
and from this the possibility of obtaining in the high-energy limit the first formulation of Very Special Relativity (VSR) follows. Every particle admits therefore its personal Maximum Attainable Velocity (MAV) as supposed by Coleman and Glashow \cite{Coleman}:
\begin{equation}
\label{a3}
c'=\left(1-f\left(\frac{|\vec{p}|}{E}\right)\right)\rightarrow(1-\epsilon)\,.
\end{equation}
This feature is present in VSR and HMSR and as a direct consequence a rich phenomenology can be obtained in different physical sectors, for instance, in neutrino oscillation physics \cite{Torrineutrini,Torri:2020dec}.

\section{Finsler geometry}
We give now a brief introduction to Finsler geometry that will be useful in the following. A Finsler geometric structure can be defined as a manifold $M$ where in every tangent space a norm function $F$ is defined not necessarily starting from an inner product. $F$ is a positively homogeneous norm only if the Hessian of $F^2$ is positive definite. The norm must be a real and positive definite function of the section of the tangent space $T_{x}M$, which depends on the point $x$ and on a vector $v\in T_{x}M$. The norm must be 1-degree homogeneous with respect to the vectors and so must satisfy the relations:
\begin{itemize}
\item $F:(M,\,T_{x}M)\rightarrow \mathbb{R}^{+}$
\item $F(x,\,v)>0\;\;\forall v\in T_{x}M,\;v\neq0,\;\forall x\in M$
\item $F(x,\,\lambda v)=|\lambda|F(x,\,v)$\,.
\end{itemize}
In Finsler geometry the norm can define a local metric via the equation:
\begin{equation}
\label{b1}
g_{\mu\nu}=\frac{1}{2}\frac{\partial^2F(x,\,v)^2}{\partial v^{\mu}\partial v^{\nu}}\,,
\end{equation}
which requires that $\det g_{\mu\nu}\neq0$ and $g_{\mu\nu}\in\mathbb{R}\;\;\forall \mu,\,\nu$. In this way it is possible to reobtain a vector norm:
\begin{equation}
\label{b2}
F(x,\,v)=\sqrt{g(x,\,v)_{\mu\nu}v^{\mu}v^{\nu}}\,.
\end{equation}
As in Riemann geometry it is possible to define a duality relation between vectors and dual forms. The definition of the Legendre transform can be posed using the metric and the resulting bijection is given by:
\begin{equation}
\label{b3}
\begin{split}
&l:T_{x}M\rightarrow T_{x}^{*}M\\
&l(v)_{\mu}=\omega_{\mu}=g(x,\,v)_{\mu\nu}v^{\nu}\,.
\end{split}
\end{equation}
The previous definition is valid for a Finsler structure with a definite signature of the metric, but it must be generalized in order to deal with the physical spacetime, which exhibits a Minkowski-type underlying metric of indefinite-signature. A systematical introduction to this issue and the so-called pseudo-Finsler geometry can be found in \cite{Pfeifer1,Pfeifer2,Pfeifer3,Hohmann,Javaloyes,Bernal}. Here we follow the definition given in \cite{Pfeifer2,Hohmann} and pose a pseudo-Finsler geometric structure modifying the previous definition and requiring that:
\begin{itemize}
\item there exists a connected component $T$ of the preimage of $F^2$  $\subset TM$, such that on $T$ the metric defined by the Hessian exists, is smooth and has Lorentzian signature $(+,\,-,\,-,\,-)$.
\end{itemize}
A typical difference between Finsler and Riemann geometry consists in the necessity to consider the fiber bundle since in the first case objects exist on $T_{x}M$ whereas in the second they do so on the manifold $M$. This fact implies that in Finsler geometry vectors and covectors must be studied in $T(TM\setminus0)$ and $T^{*}(TM\setminus0)$, respectively. Moreover, it is necessary to introduce the horizontal-vertical decomposition and the related nonlinear connection for the $T(TM\setminus0)$ and $T^{*}(TM\setminus0)$ structures, as in \cite{miron2012lagrangian}.\\
Considering $T^{*}(TM\setminus0)$ one can introduce the \emph{vertical distribution} generated by the derivative $\partial/\partial v$ as:
\begin{equation}
\label{b4}
V:w\in T^{*}M\rightarrow V_{w}\subset T_{w}(T^{*}M)\,.
\end{equation}
Given the vertical distribution, it is possible to identify a complementary structure, the \emph{horizontal distribution}, named \emph{nonlinear connection}, for which the Whitney decomposition is valid \cite{Vacaru}:
\begin{equation}
\label{b5}
T_{w}(T^{*}M)=N_{w}\oplus V_{w}\,.
\end{equation}
The nonlinear connection is a collection of homogeneous functions of degree 1, locally defined on the manifold, such that:
\begin{equation}
\label{b6}
\frac{\delta}{\delta x^{\mu}}=\frac{\partial}{\partial x^{\mu}}+N_{\mu\nu}(x,\,p)\frac{\partial}{\partial p_{\nu}}\,.
\end{equation}
In this work we follow \cite{miron2012lagrangian} and as nonlinear connection we choose the usual General Relativity (GR) form:
\begin{equation}
\label{b7}
N_{\mu\nu}(x,\,p)=\Gamma_{\mu\nu}^{\alpha}(x)p_{\alpha}\,,
\end{equation}
where $\Gamma_{\mu\nu}^{\alpha}$ represents the GR affine connection. The choice for a nonlinear connection is not uniquely defined, but this particular formulation is the most convenient one in this context.\\
Analogously, the cotangent space is spanned by the differential basis defined by:
\begin{equation}
\label{b8}
\delta p_{\mu}=\mathrm{d}p_{\mu}-N_{\mu\nu}(x,\,p)\mathrm{d}x^{\nu}\,.
\end{equation}
In the context of Finsler geometry it is possible to generalize the definition of the affine connection via the general Christoffel symbols:
\begin{equation}
\label{b8}
H_{\mu\nu}^{\alpha}(x,\,p)=\frac{1}{2}g^{\alpha\beta}(x,\,p)\left(\frac{\delta g_{\beta,\nu}(x,\,p)}{\delta x^{\mu}}+\frac{\delta g_{\mu,\beta}(x,\,p)}{\delta x^{\nu}}-\frac{\delta g_{\mu,\nu}(x,\,p)}{\delta x^{\beta}}\right)
\end{equation}
and the geodesic equations become:
\begin{equation}
\label{b9}
\frac{\mathrm{d}^2x^{\alpha}}{\mathrm{d}\tau^2}+H_{\mu\nu}^{\alpha}\frac{\mathrm{d}x^{\mu}}{\mathrm{d}\tau}\frac{\mathrm{d}x^{\nu}}{\mathrm{d}\tau}=0\,.
\end{equation}
The study conducted in this work can be generalized with these prescriptions in a curved spacetime. In the following, we will consider a geometry with no matter induced curvature for the investigation of the phenomenology introduced in the propagation of UHECRs. Hence, the nonlinear connection \cref{b7} is zero and the derivative \cref{b6} reduces to the usual partial derivative.

\section{Finsler geometry and HMSR}
The MDR (\ref{a1}) can be interpreted as the Finsler pseudo-norm characterizing the spacetime (momentum space) structure.
In HMSR the standard physics is modified in a CPT-even scenario, the MDRs, indeed, do not exhibit any dependence on helicities, for instance, and are supposed to be equal for particles and the related antiparticles. A well-known result about fundamental physics symmetry states that in a CPT-odd scenario the Lorentz symmetry as usually formulated must be violated, whereas the opposite statement is not automatically true and therefore it is possible to violate or modify covariance, while preserving the CPT symmetry \cite{Greenberg1,Greenberg2}. A review on the implications of the CPT symmetry in particle and astroparticle physics and its relation to the Lorentz/Poincaré symmetry is presented, for instance, in \cite{TorriCPT}.\\
Promoting the MDR \cref{a1} to the role of a Finsler norm, it is possible to obtain the related metric of the momentum space as the Hessian of the squared norm:
\begin{equation}
\label{a4}
\widetilde{g}^{\mu\nu}(p)=\frac{1}{2}\frac{\partial}{\partial p_{\mu}}\frac{\partial}{\partial p_{\nu}}F^2(E,\,\vec{p})\,.
\end{equation}
Since the function $F^2$ is 2-degree homogeneous, the computation produces a 0-homogeneous tensor that can be written as:
\begin{equation}
\label{a5}
\widetilde{g}^{\mu\nu}(p)=D_{4\times4}+A_{4\times4}\,,
\end{equation}
with a diagonal matrix $D$:
\begin{equation}
\label{a5a}
D=\left(
      \begin{array}{cc}
            \left(1+h\left(\left|\vec{p}\right|/E\right)\right) & \vec{0} \\
            \vec{0}^{t} & -\left(1+k\left(\left|\vec{p}\right|/E\right)\right)\mathbb{I}_{3\times3} \\
      \end{array}
\right)\,.
\end{equation}
The matrix $D$ is the associated generalized Hamilton's space metric that effectively contributes to the computation of the MDR \cref{a1}. Indeed, by direct calculation it is straightforward to demonstrate the relation:
\begin{equation}
\label{a5aa}
p_{\mu}\widetilde{g}^{\mu\nu}(p)p_{\nu}=p_{\mu}D^{\mu\nu}(p)p_{\nu}=E^2\left(1+h\left(\frac{\left|\vec{p}\right|}{E}\right)\right)-\left|\vec{p}\right|^2\left(1+k\left(\frac{\left|\vec{p}\right|}{E}\right)\right)\,.
\end{equation}
The complete Finsler metric tensor is written including the matrix $A$ \cref{a5} and the metric tensor defined in this way satisfies the Finsler requirement of a totally symmetric associated Cartan's tensor:
\begin{equation}
\label{a5b}
C_{\mu\nu\alpha}=\frac{1}{2}\frac{\partial^{3}F^2(p)}{\partial p_{\mu} \partial p_{\nu} \partial p_{\alpha}}\,.
\end{equation}
The matrix $A$ has both diagonal and nondiagonal entries and thanks to the explicit form acquired by the metric tensor $\widetilde{g}$ it is simple to find out that this matrix gives no contribution to the norm of a covector: $p_{\mu}A^{\mu\nu}p_{\nu}=0$ \cref{a5aa}. For the following computations it is important to emphasize that the entries of this matrix are first-order perturbations under the assumption that the functions $h,\,k\simeq O(1)$ and their derivatives have a perturbative character\footnote{See Appendix A}, hence $A\simeq O(1)$.\\
From the Finsler co-metric one can derive the metric associated with coordinate space as the inverse one, via the defining equation:
\begin{equation}
\label{a6}
\widetilde{g}^{\mu\alpha}g_{\alpha\nu}=\delta^{\mu}_{\,\nu}\,,
\end{equation}
obtaining the explicit form of the coordinate space metric:
\begin{equation}
\label{a6}
g_{\mu\nu}(p)=D^{*}-A_{4\times4}+O(2)\\
\end{equation}
and the diagonal matrix $D^{*}$:
\begin{equation}
\label{a6a}
D^{*}=\left(
          \begin{array}{cc}
             \left(1-h\left(\left|\vec{p}\right|/E\right)\right) & \vec{0} \\
             \vec{0}^{t} & -\left(1-k\left(\left|\vec{p}\right|/E\right)\right)\mathbb{I}_{3\times3} \\
          \end{array}
\right)\,,
\end{equation}
where $O(2)$ indicates a second-order perturbation.

\subsection{Hamilton and Lagrangian Finsler geometry}
The correlation between coordinate and momentum space must be further explored and it can be done starting from the action:
\begin{align}
\label{a7}
S&=\int \mathcal{L}\,\mathrm{d}\tau=\int \left(\dot{x}^{\mu}p_{\mu}-\frac{\lambda}{2} F^{2}(p)\right)\mathrm{d}\tau \notag \\
&=\int \left(\dot{x}^{\mu}p_{\mu}-\frac{\lambda}{2} \left(\widetilde{g}^{\alpha\beta}(p)p_{\alpha}p_{\beta}-m^2\right)\right)\mathrm{d}\tau\,,
\end{align}
where the Lagrangian $\mathcal{L}$ is defined using a Lagrangian multiplier $\lambda$ to pose the constraint that the MDR, $F^{2}(p)=m^2$, must be satisfied \cite{Liberati1}. The action $S$ is obtained integrating the Lagrangian $\mathcal{L}$ with respect to the proper time $\tau$.\\
The equations of motion can be obtained varying the Lagrangian $\mathcal{L}$. Hence, one can obtain the partial derivatives with respect to the velocity $\dot{x}^{\mu}$ and the Lagrange multiplier $\lambda$:
\begin{subequations}
\label{a8}
\begin{align}
\frac{\partial\mathcal{L}}{\partial \dot{x}^{\mu}}&=0\;\Rightarrow\;\frac{\mathrm{d} p_{\mu}}{\mathrm{d} \tau}=0\\
\frac{\partial\mathcal{L}}{\partial \lambda}&=0\;\Rightarrow\;\widetilde{g}^{\mu\nu}(p)p_{\mu}p_{\nu}=m^2\,.
\end{align}
\end{subequations}
The variation calculated with respect to the momentum $p_{\mu}$ gives:
\begin{equation}
\label{a9}
\frac{\partial\mathcal{L}}{\partial p_{\mu}}=0\;\Rightarrow\;\dot{x}^{\mu}=\lambda \widetilde{g}^{\mu\nu}(p)p_{\nu}+\frac{\lambda}{2}\left(\frac{\partial \widetilde{g}^{\alpha\beta}(p)}{\partial p_{\mu}}\right)p_{\alpha}p_{\beta}=\,.
\end{equation}
A simple application of the Euler's theorem on homogeneous functions implies that the second term on the right-hand side of the previous relation is zero thanks to the 0-homogeneity of the metric tensor. Therefore, the velocity can be written in the simple form:
\begin{equation}
\label{a9a}
\dot{x}^{\mu}=\lambda \widetilde{g}^{\mu\nu}(p)p_{\nu}
\end{equation}
Inverting the previous relation, the momentum can be expressed as:
\begin{equation}
\label{a10}
p_{\mu}=\lambda^{-1}g_{\mu\nu}\dot{x}^{\nu}\,.
\end{equation}
Substituting the previous relation in the action \cref{a7} it is possible to compute the Hamiltonian:
\begin{equation}
\label{a10a}
\mathcal{H}=\frac{\lambda}{2}\widetilde{g}^{\mu\nu}p_{\mu}p_{\nu}+\frac{\lambda}{2}m^{2}\,,
\end{equation}
which acquires this simple explicit form since the second term on the right-hand side in \cref{a10} is orthogonal to the momentum $p_{\mu}$.\\
The momentum can be expressed as a function of the velocity solving \cref{a10}, for instance, perturbatively. As a consequence, the coordinate space metric $\widetilde{g}_{\mu\nu}(p)$, defined as the inverse of the momentum space metric $\widetilde{g}^{\mu\nu}(p)$, can be written as a function of coordinate $x$ and velocity $\dot{x}$:
\begin{equation}
\label{a15x}
\widetilde{g}_{\mu\nu}(p)=g_{\mu\nu}(\dot{x}).
\end{equation}
The Lagrangian can be computed substituting \cref{a9} in the action \cref{a7}:
\begin{equation}
\label{a12}
\mathcal{L}=\frac{1}{2\lambda}g_{\mu\nu}(p)\dot{x}^{\mu}\dot{x}^{\nu}+\frac{\lambda}{2}m^2\,,
\end{equation}
where $O(2)$ is a second-order perturbation with respect to the scale fixed by the functions $h$ and $k$ (\ref{a1a}) again thanks to the orthogonality of $p_{\mu}$ and the second term on the right-hand side of \cref{a10}.
The variation of the previous relation with respect to the parameter $\lambda$ gives:
\begin{equation}
\label{a13x}
\frac{\partial\mathcal{L}}{\partial\lambda}=0\;\Rightarrow\;\frac{g_{\mu\nu}(\dot{x})\dot{x}^{\mu}\dot{x}^{\nu}}{\lambda^2}=m^2\;\Rightarrow\;\lambda=\frac{\sqrt{g_{\mu\nu}(\dot{x})\dot{x}^{\mu}\dot{x}^{\nu}}}{m}
\end{equation}
and then the Lagrangian can be written as:
\begin{equation}
\label{a14x}
\mathcal{L}=m\sqrt{{g_{\mu\nu}(\dot{x})\dot{x}^{\mu}\dot{x}^{\nu}}}\,.
\end{equation}
The relation between the Lagrangian and the Hamiltonian formulation has been established and it is possible to compute the norm for the velocity $\dot{x}$. This function is associated with the momentum space norm $F$ which is defined via the MDR \cref{a1}. The coordinate space norm can be determined starting from the metric \cref{a6}:
\begin{align}
\label{a15ax}
F(\dot{x})^2&=\dot{x}^{\mu}g_{\mu\nu}(\dot{x})\dot{x}^{\nu}=\dot{x}^{\mu}\left(D^{*}_{\mu\nu}-A_{\mu\nu}+O(2)\right)\dot{x}^{\nu} \notag \\
&=\left(1-h\left(\left|\vec{p}\right|/E\right)\right)\dot{x}_{0}^{2}-\left(1-k\left(\left|\vec{p}\right|/E\right)\right)\vec{x}^{2}+O(2)\,.
\end{align}
As a final remark it is possible to state that the resulting structure is a Finsler geometry.

\section{HMSR generalized Finsler spacetime}
In the following we will consider a generalized Finsler geometry model to set the stage of the HMSR formulation. In this context we will obtain an extension of the Standard Model (SM) of particle physics and will conduct the computations related with the GZK cutoff phenomenon.\\
The generalized Finsler geometry is a less restrictive structure which does not require a totally symmetric Cartan's tensor associated to the metric \cref{a5b}, therefore, the construction of the geometry is easier. In this context the spacetime and the momentum metrics \cref{a5,a6} can be simplified taking into account only the diagonal parts \cref{a5a,a6a}:
\begin{align}
&g_{\mu\nu}(p)=D=
      \left(
      \begin{array}{cc}
            \left(1+h\left(\left|\vec{p}\right|/E\right)\right) & \vec{0} \\
            \vec{0}^{t} & -\left(1+k\left(\left|\vec{p}\right|/E\right)\right)\mathbb{I}_{3\times3} \\
      \end{array}
      \right)\\
&g^{\mu\nu}(p)=D^{*}=
       \left(
       \begin{array}{cc}
             \left(1-h\left(\left|\vec{p}\right|/E\right)\right) & \vec{0} \\
             \vec{0}^{t} & -\left(1-k\left(\left|\vec{p}\right|/E\right)\right)\mathbb{I}_{3\times3} \\
       \end{array}
       \right)\,.
\end{align}
In this way the coordinate and momentum metrics are reduced to the parts that really contribute in evaluating the squared norm of a vector or a covector respectively.\\
Starting from the defining equations:
\begin{subequations}
\label{a13}
\begin{align}
g_{\mu\nu}(p)&=e_{\mu}^{\,a}(p)\,\eta_{ab}\,e_{\nu}^{\,b}(p)\\[1ex]
\widetilde{g}^{\mu\nu}(p)&=e^{\mu}_{\,a}(p)\,\eta^{ab}\,e^{\nu}_{\,b}(p)\,.
\end{align}
\end{subequations}
it is possible to compute the generalized associated vierbein, which can be written as:
\begin{subequations}
\label{a14}
\begin{align}
e_{\mu}^{\,a}(p)&=
\left(
    \begin{array}{cc}
        \sqrt{1-h\left(\left(\left|\vec{p}\right|/E\right)\right)} & \vec{0} \\
        \vec{0}^{t} & \sqrt{1-k\left(\left(\left|\vec{p}\right|/E\right)\right)}\,\mathbb{I}_{3\times3} \\
    \end{array}
\right)\\[1ex]
e_{\,a}^{\mu}(p)&=
\left(
    \begin{array}{cc}
        \sqrt{1+h\left(\left(\left|\vec{p}\right|/E\right)\right)} & \vec{0} \\
        \vec{0}^{t} & \sqrt{1+k\left(\left(\left|\vec{p}\right|/E\right)\right)}\,\mathbb{I}_{3\times3} \\
    \end{array}
\right)\,.
\end{align}
\end{subequations}
It is simple to determine the perturbative order $O(1)$ of part of these matrix entries
using the approximation $\sqrt{1\pm\epsilon}\simeq1\pm\frac{1}{2}\epsilon$, valid for $\epsilon\ll1$.\\
The form of the vierbein \cref{a14} will be used in all the following computations, for instance, when evaluating the kinematical invariants (the Mandelstam $s$ variable of the reactions involved in the GZK cut-off phenomenon).\\

\subsection{Generalized covariance}
In the HMSR model every particle has its personal modified spacetime, which is parameterized by the particle momentum. This means that every physical quantity related with a given particle is generalized and lives in a spacetime that acquires an explicit dependence on the particle energy. It is therefore necessary to introduce an original formalism to correlate different local spaces, using the generalized vierbein elements as projectors from every spacetime to a common flat Minkowski support space. Here we report a scheme of how the correlation between different local spaces is established:
\[\begin{tikzcd}
         (TM,\,\eta_{ab},\,p) \arrow{d}{e(\mathbf{p})} \arrow{rr}[swap]{\Lambda} && (TM,\,\eta_{ab},\,p')\arrow{d}[swap]{\overline{e}(p')} \\
        (T_{x}M,\,g_{\mu\nu}(p)) \arrow{rr}[swap]{\overline{e}\circ\Lambda\circ e^{-1}} && (T_{x}M,\,\overline{g}_{\mu\nu}(p'))\,.
\end{tikzcd}\]
The Greek indices refer to the local curved geometric structures, whereas the Latin ones refer to the common Minkowski support space. Referring to the previous scheme, using the vierbein elements as projectors, the generalized Lorentz transformations can be obtained:
\begin{equation}
\label{a15}
\Lambda_{\mu}^{\;\nu}(p)=e_{\;\mu}^{a}(\Lambda p)\,\Lambda_{a}^{\;b}\,e_{\;\nu}^{b}(p)\,,
\end{equation}
where the elements $\Lambda_{a}^{\;b}$ belong to the Lorentz group and are defined in the flat Minkowski spacetime and the used vierbein is defined in \cref{a14} and includes all terms up to the perturbative order under consideration. The Lorentz covariance is promoted to a diffeomorphism invariance and the introduced class of \emph{Modified Lorentz Transformations} (MLT) represent the isometries of the MDR of \cref{a1}.\\
We point out that the previous construction about the generalization of covariance is valid if the spacetime is metrizable and admits a vierbein, hence the result is valid even in different contexts. Therefore, the present prescription can be used to pose analogous definitions and set the stage of the model in the context of the more restrictive Finsler geometry.

\subsection{Affine and spinorial connections}
The geometrical structure is characterized by the \emph{affine} and the \emph{spinorial} connections. In Finsler geometry the affine connection, that is the Christoffel symbol, can be defined using \cref{b8}. In the case of zero spacetime curvature, the nonlinear connection \cref{b7} is $N_{\mu\nu}=0$ and therefore the derivative \cref{b8} reduces to the ordinary one:
\begin{equation}
\label{a16}
\frac{\delta}{\delta x^{\mu}}=\frac{\partial}{\partial x^{\mu}}\,.
\end{equation}
Since in the case of absence of curvature, the metric depends only on the momentum (or equivalently on the velocity) and not on the coordinates, the Christoffel symbol \cref{b8} becomes:
\begin{equation}
\label{a17}
H_{\mu\nu}^{\alpha}(x,\,p)=\frac{1}{2}g^{\alpha\beta}(p)\left(\frac{\partial g_{\beta,\nu}(p)}{\partial x^{\mu}}+\frac{\partial g_{\mu,\beta}(p)}{\partial x^{\nu}}-\frac{\partial g_{\mu,\nu}(p)}{\partial x^{\beta}}\right)=0\,.
\end{equation}
The explicit form of the covariant derivative is a consequence of this result and it is equal to the ordinary derivative in flat spacetime:
\begin{equation}
\label{a18}
\nabla_{\mu}v^{\nu}=\partial_{\mu}v^{\nu}+H_{\mu\alpha}^{\nu}v^{\alpha}=\partial_{\mu}v^{\nu}\,.
\end{equation}
Now it is possible to define the spinorial connection using the covariant derivative:
\begin{equation}
\label{a19}
\omega_{\mu ab}=e_{a}^{\,\nu}\nabla_{\mu}e_{b\nu}=e_{a}^{\,\nu}\partial_{\mu}e_{b\nu}\,.
\end{equation}
It is simple to demonstrate that all the connection coefficients vanish since the vierbein depends on the momentum but not on the coordinates. Finally we obtain the total geometric covariant derivative, which will be useful in defining the minimal extension of the Standard Model (SM) of particle physics:
\begin{equation}
\label{a22}
D_{\mu}v^{\;\nu}_{a}=\partial_{\mu}v^{\;\nu}_{a}+\Gamma_{\mu\alpha}^{\,\nu}v^{\;\alpha}_{a}-\omega_{\mu\nu}^{\,a}v^{\;\nu}_{b}\simeq\partial_{\mu}v^{\;\nu}_{b}\,.
\end{equation}
The resulting spacetime geometry is therefore a flat Finsler pseudo-structure \cite{Koste11,Liberati1,Edwards,Lammerzahl,Bubuianu,Schreck}, whereas the associated momentum space is asymptotically flat.

\subsection{Modified Poincaré brackets}
As a final result about the geometrical structure introduced by HMSR, we report the modified Poincaré brackets, computed for the local structure using the vierbein projectors:
\begin{subequations}
\begin{align}
\label{a23}
\{\widetilde{x}^{\mu},\,\widetilde{x}^{\nu}\}&=\{x^{i}e_{i}^{\,\mu}(p),\,x^{j}e_{j}^{\,\nu}(p)\}=\{x^{i},\,e_{j}^{\,\nu}(p)\}e_{i}^{\,\mu}(p)x^{j}+\{e_{i}^{\,\mu}(p),\,x^{j}\}x^{i}e_{j}^{\,\nu}(p)\\
\{\widetilde{x}^{\mu},\,\widetilde{p}_{\nu}\}&=\{x^{i}e_{i}^{\,\mu}(p),\,p_{j}e^{j}_{\,\nu}(p)\}=\{x^{i},\,p_{j}\}e_{i}^{\,\mu}(p)e^{j}_{\,\nu}(p)+\{x^{i},\,e^{j}_{\,\nu}(p)\}e_{i}^{\,\mu}(p)p_{j} \notag \\
&=\delta^{\mu}_{\,\nu}+\{x^{i},\,e^{j}_{\,\nu}(p)\}e_{i}^{\,\mu}(p)p_{j}\\
\{\widetilde{p}_{\mu},\,\widetilde{p}_{\nu}\}&=\{p_{i}e^{i}_{\,\mu}(p),\,p_{j}e^{j}_{\,\nu}(p)\}=0\,,
\end{align}
\end{subequations}
where $\{\widetilde{x}^{\mu}\}$ are the local coordinates and $\{x^{i},\,e^{j}_{\,\nu}(p)\}\neq0$ since the vierbein $e^{j}_{\,\nu}(p)$ is a function of the momentum $p_{\mu}$. Hence, time and space coordinates do not commute anymore \cite{Torri:2020dec}. Using the explicit form of the vierbein \cref{a14} and admitting for the perturbation functions the approximations\footnote{See Appendix A}:
\begin{subequations}
\begin{align}
h\left(\frac{|\vec{p}|}{E}\right)=\alpha\left(\frac{|\vec{p}|}{E}\right)+\dots\\
k\left(\frac{|\vec{p}|}{E}\right)=\beta\left(\frac{|\vec{p}|}{E}\right)+\dots
\end{align}
\end{subequations}
it is possible to obtain the following relations valid at the first perturbative order:
\begin{subequations}
\begin{align}
\label{a23}
[\widetilde{x}^{\mu},\,\widetilde{x}^{\nu}]&=\theta^{\mu\nu}\\
[\widetilde{x}^{\mu},\,\widetilde{p}_{\nu}]&=\delta^{\mu}_{\,\nu}+[x^{i},\,e^{j}_{\,\nu}(p)]e_{i}^{\,\mu}(p)p_{j}\simeq\delta^{\mu}_{\,\nu}-\frac{1}{2}(\alpha+3\beta)\left(\frac{|\vec{p}|}{E}\right)^2\\
[\widetilde{p}_{\mu},\,\widetilde{p}_{\nu}]&=0\,,
\end{align}
\end{subequations}
with the antisymmetric matrix $\theta^{\mu\nu}$ that satisfies the relations:
\begin{align}
&\theta^{ij}=\theta^{00}=0\quad \forall i,j\in{1,\,2,\,3}\\
&\theta^{0i}=-\theta^{i0}\simeq\frac{1}{2}\left(\beta\left(\frac{|\vec{p}|x^{i}}{E^2}\right)-\alpha\left(\frac{|\vec{p}|x^{0}}{E|\vec{p}|}\right)\right)\simeq\frac{1}{2}\left(\beta-\alpha\right)
\end{align}
in the high energy limit, proving that, if the parameters $\alpha$ and $\beta$ are different, the spacetime coordinates do not commute anymore \cite{Seiberg}.\\
The relation existing between noncommutative field theory and LIV was analyzed in \cite{Carroll}; here we introduced a framework to investigate this point in the context of modified covariance.

\section{Minimal extension of the Standard Model in an isotropic scenario}
In the HMSR framework the SM of particle physics can be amended in order to include the QG-caused perturbations preserving covariance of the theory, even if in an amended formulation. Following, for instance, a strategy analogous to that used for the isotropic sector of the \emph{Standard Model Extension} (SME) \cite{Koste1}, the theory formulation requires the definition of the modified Dirac matrices with the related Clifford algebra and spinors.

\subsection{Modified Clifford algebra and spinors}
The Dirac matrices acquire an explicit dependence on the particle momentum and again using the vierbein projectors \cref{a14} they can be written as:
\begin{equation}
\label{a24}
\Gamma^{\mu}=e^{\;\mu}_{a}(p)\,\gamma^{a}\qquad \Gamma_5=\frac{\epsilon^{\mu\nu\alpha\beta}}{4!}\Gamma_{\mu}\Gamma_{\nu}\Gamma_{\alpha}\Gamma_{\beta}=\gamma_5\,.
\end{equation}
The $\gamma_{5}$ matrix is constant and this means that the chiral projectors are not affected.\\
The Dirac matrices modified via \cref{a24} satisfy the defining relation of the Clifford algebra:
\begin{equation}
\label{a25}
\{\Gamma_{\mu},\Gamma_{\nu}\}=2\,g^{\mu\nu}(p)=2\,e_{\mu}^{\,a}(p)\,\eta_{ab}\,e_{\nu}^{\,b}(p)\,.
\end{equation}
The definition of spinor fields is now amended preserving the usual plane-wave formulation:
\begin{subequations}
\label{a26}
\begin{align}
\psi^{+}(x)&=u_{r}(p)\mathrm{e}^{-\mathrm{i}p_{\mu}x^{\mu}}\\
\psi^{-}(x)&=v_{r}(p)\mathrm{e}^{\mathrm{i}p_{\mu}x^{\mu}}\,.
\end{align}
\end{subequations}
The normalization of the spinors $u_{r}(p)$ and $v_{r}(p)$ is modified since these are defined using the newly introduced metric \cref{a6} and the related internal product.\\
From the previous definitions the modified Dirac equation can be derived:
\begin{equation}
\label{a27}
\left(i\Gamma^{\mu}\partial_{\mu}-m\right)\psi=0\,.
\end{equation}
An important consistency check of the new formalism can be obtained verifying that \cref{a27} implies the MDR \cref{a1}.
\begin{subequations}
\label{a28}
\begin{align}
&(i\Gamma^{\mu}\partial_{\mu}+m)(i\Gamma^{\mu}\partial_{\mu}-m)\psi^{+}=0 \\[1ex]
&\Rightarrow (\Gamma^{\mu}p_{\mu}+m)(\Gamma^{\mu}p_{\mu}-m)u_{r}(p)=0 \\[1ex]
&\Rightarrow \left(\frac{1}{2}\{\Gamma^{\mu},\,\Gamma^{\nu}\}p_{\mu}p_{\nu}-m^2\right)u_{r}(p)=0 \\[1ex]
&\Rightarrow (p_{\mu}g^{\mu\nu}p_{\nu}-m^2)u_{r}(p)=0\,.
\end{align}
\end{subequations}
As a final result, a minimal extension of the SM can be obtained from the formalism here introduced. Indeed using the vierbein elements to project to a common Minkowski support spacetime the physical quantities related to different interacting particles and using the explicit form of the total covariant derivative \cref{a22}, the minimal extension of the SM can be formulated for a flat spacetime. Here we illustrate the amended formulation of quantum electrodynamics (QED), whose Lagrangian can be written in the form:
\begin{equation}
\label{a29}
\mathcal{L}=\sqrt{\left|\det{[g]}\right|}\;\;\overline{\psi}(i\Gamma^{\mu}\partial_{\mu}-m)\psi+e\sqrt{\left|\det{[\widetilde{g}]}\right|}\;\;\overline{\psi}\,\Gamma_{\mu}(p,\,p')\,\psi\,\overline{e}^{\mu}_{\;a}\,A^{a}\,,
\end{equation}
where the term $\sqrt{|\det{[g]}|}$ is borrowed from the formulation of QFT in curved spacetime. The vierbein element $\overline{e}$ is related to the gauge field $A^{a}(x)$. The gauge field is supposed as Lorentz-covariant in the usual meaning, that is, the MAV of photons is the usual speed of light $c$. The gauge field therefore is set on a Minkowski spacetime $(TM,\,\eta_{ab})$ and the vierbein is given by: $\overline{e}^{\mu}_{\;a}=\delta^{\mu}_{\;a}$. The QG corrections can be introduced in the generic gauge boson sector modifying the definition of the vierbein related to the gauge field, as done for the massive fermion fields.\\
The interaction is governed by the conserved current.
\begin{equation}
\label{a30}
J_{\mu}=e\sqrt{\left|\det{\tfrac{1}{2}\{\Gamma_{\mu},\,\Gamma_{\nu}\}}\right|}\;\;\overline{\psi}\,\Gamma_{\mu}\,\psi=e\sqrt{\left|\det{[g]}\right|}\;\;\overline{\psi}\,\Gamma_{\mu}\,\psi\,.
\end{equation}
In the low-energy scenario the covariance perturbations in the conserved current are negligible, whereas in the high-energy limit the formulation admits a constant form, since one can consider the incoming and outgoing momenta with the same constant high-energy limit.

\subsection{Gauge symmetry}
The SM minimal extension obtained in the context of HMSR preserves the classic internal gauge symmetry $SU(3)\otimes SU(2)\otimes U(1)$. This result can be stated formulating an amended version of the Coleman-Mandula theorem \cite{TorriHMSR}. The modified gauge symmetry group acquires the explicit form:
\begin{equation}
\label{a31}
\mathcal{P}(\{p\})\otimes G_{int}\,,
\end{equation}
where $\mathcal{P}(\{p\})$ is the kinematical symmetry group and is given by the direct product of the momentum-dependent Poincaré groups associated with the different particle species
\begin{equation}
\label{a32}
\mathcal{P}\left(\{p\}\right)=\bigotimes_{i}\mathcal{P}^{(i)}\bigr(p_{(i)}\bigr)
\end{equation}
and $G_{\mathrm{int}}$ is the internal gauge symmetry group, in this case the usual SM gauge group:
\begin{equation}
\label{a33}
G_{\mathrm{int}}=SU(3)\otimes SU(2)\otimes U(1)\,.
\end{equation}

\subsection{Modified kinematics}
HMSR theory perturbs the kinematics geometrizing the interaction of free propagating particles with the supposed quantized structure of the spacetime. In this work we investigate the phenomenological effects introduced by QG in the UHECRs propagation, where the main effects are caused by the modification of the kinematics. The detectable effects are caused by the interaction of different particle species, which modify in a proper way the related spacetime.\\
Introducing a generalized internal product in the momentum space for more than two different interacting particle species, it is possible to obtain a generalized formulation for the Mandelstam variables \emph{s}, \emph{t} and \emph{u}, which are the dynamical invariant quantities linked to a reaction.\\
Considering $p$ and $q$ as the momenta of two interacting particles of different species, the internal product of their sum can be defined as:
\begin{equation}
\label{a34}
\big( p+q|p+q \big)=
(p_{\mu}\,e_{a}^{\,\mu}(p)+q_{\mu}\,\tilde{e}_{a}^{\,\mu}(q))\,\eta^{ab}\,(p_{\nu}\,e_{b}^{\,\nu}(p)+q_{\nu}\,\tilde{e}_{b}^{\,\nu}(q))\,,
\end{equation}
where the vierbeins $e_{a}^{\,\mu}(p)$ and $\tilde{e}_{a}^{\,\mu}(q))$ are associated, respectively, with the two different particle species. The formulation of the modified internal product can be simplified in the form:
\begin{align}
\label{a35}
\big( p+q|p+q \big)&=
\begin{pmatrix}
    p \\
    q \\
  \end{pmatrix}
^{t}\cdot G\cdot
  \begin{pmatrix}
    p \\
    q \\
  \end{pmatrix}=
\begin{pmatrix}
    p & q \\
  \end{pmatrix}
\left(
    \begin{array}{cc}
      g^{\mu\nu}(p) & e^{a\mu}(p)\tilde{e}_{a}^{\;\beta}(q) \\
      \tilde{e}^{a\alpha}(q)e_{a}^{\;\nu}(p) & \tilde{g}^{\alpha\beta}(q) \\
    \end{array}
\right)
  \begin{pmatrix}
    p \\
    q \\
  \end{pmatrix} \notag \\
&=p_{\mu}\,g^{\mu\nu}(p)\,p_{\nu}+p_{\mu}\, e^{a\mu}\tilde{e}_{a}^{\;\beta}(q)\,q_{\beta} \notag \\
&\phantom{{}={}}+q_{\alpha}\,\tilde{e}^{a\alpha}(q)\,e_{a}^{\;\nu}(p)\,p_{\nu}+q_{\alpha}\tilde{g}^{\alpha\beta}(q)\,q_{\beta}\,,
\end{align}
using the generalized metric:
\begin{equation}
\label{a36}
G=\left(
    \begin{array}{cc}
      g^{\mu\nu}(p) & e^{a\mu}(p)\tilde{e}_{a}^{\;\beta}(q) \\
      \tilde{e}^{a\alpha}(q)e_{a}^{\;\nu}(p) & \tilde{g}^{\alpha\beta}(q) \\
    \end{array}
  \right)\,.
\end{equation}
The inner product defined in \cref{a35} is invariant with respect to the \emph{Modified Lorentz Transformations} (MLT) introduced in HMSR:
\begin{equation}
\label{a37}
\Lambda=
\left(
  \begin{array}{cc}
    \Lambda_{\mu}^{\;\mu'} & 0 \\
    0 & \tilde{\Lambda}_{\alpha}^{\;\alpha'} \\
  \end{array}
\right)\,.
\end{equation}
The inner product remains invariant under the action of such generalized Lorentz transformations, indeed one can obtain:
\begin{align}
\label{a38}
\big( p+q|p+q \big)&=
  \begin{pmatrix}
    p \\
    q \\
  \end{pmatrix}
^{t}\cdot G\cdot
  \begin{pmatrix}
    p \\
    q \\
  \end{pmatrix}=\left(\Lambda
  \begin{pmatrix}
    p \\
    q \\
  \end{pmatrix}
\right)^{t}\cdot \Lambda \cdot G\cdot \Lambda^{t}\cdot\Lambda
  \begin{pmatrix}
    p \\
    q \\
  \end{pmatrix} \notag \\
&=\big( \Lambda(p+q)|\Lambda(p+q) \big)\,.
\end{align}
$\Lambda \cdot G\cdot \Lambda^{t}$ is the metric evaluated for the two particle momenta $\Lambda p$, $\tilde\Lambda q$.\\
The new formalism here introduced guarantees that HMSR theory can deal with the interaction of different particle species in the context of a QG-modified kinematics without the necessity of the introduction of a preferred reference frame.

\section{Ultra-high-energy cosmic-ray propagation}
Before investigating the QG-induced phenomenology in UHECRs physics, it is useful to illustrate the standard physics predictions for these highly energetic particles. During their propagation, UHECRs can interact with the CMB and depending on their nature and energy are attenuated. For instance, a propagating CR bare nucleus $N$ with atomic number $A$ can undergo a photo-dissociation process interacting with the CMB:
\begin{equation}
\label{a39}
N_{A}+\gamma\,\rightarrow\,N_{(A-1)}+n\,,
\end{equation}
where $n$ represents a generic nucleon. The proton propagation is influenced by different interaction processes with the CMB, hence it can undergo, for instance, a pair production process:
\begin{equation}
\label{a40}
p+\gamma\,\rightarrow\,p+e^{-}+e^{+}\,.
\end{equation}
This process is the main interaction mechanism with the CMB for CR protons with an energy lower than a threshold $E\sim \unit[5\times10^{19}]{eV}$. The dominant process for UHECR protons with an energy exceeding this threshold is the $\Delta$ particle resonance photopion production process:
\begin{equation}
\label{a41}
\begin{split}
&p+\gamma\,\rightarrow\,\Delta\,\rightarrow\,p+\pi^{0}\\
&p+\gamma\,\rightarrow\,\Delta\,\rightarrow\,n+\pi^{+}\,.
\end{split}
\end{equation}
These dissipation mechanisms make the Universe opaque to the propagation of CR, particularly for the most energetic component (UHECR), with an energy that exceeds the threshold $E\sim \unit[5\times10^{19}]{eV}$. In this work we are particularly interested in the UHECRs propagation since CRs can be useful in investigating QG phenomenology thanks to their huge energy and propagation length. QG effects in some theories are expected as more evident at high energies and their perturbations can sum up during the propagation of this kind of particles. The photopion production is the main attenuation interaction of UHECRs and is the core mechanism for the so-called GZK cut-off phenomenon for protons \cite{Greisen,Zatsepin}. This effect poses an upper limit on the energy of protons detected at ground and coming from distant sources. Since through this effect a particle dissipates energy but is not annihilated, a proton with enough energy can undergo the same interaction process again and can undergo what is a stochastic dissipation process. This way, it becomes possible to evaluate the \emph{attenuation length}, defined as the average distance that the proton has to travel in order to reduce its energy by a factor of $1/\mathrm{e}$. The inverse of the attenuation length is given by \cite{Stecker1}:
\begin{align}
\label{a42}
\frac{1}{l_{p\gamma}}&=\int_{\epsilon_{\mathrm{th}}}^{+\infty}n(\epsilon)\,\mathrm{d}\epsilon\int_{-1}^{+1}\frac{1}{2}\,s\,(1-\mu)\,\sigma_{p\gamma}(s)\,K(s)\,\mathrm{d}\mu \\
&=\int_{\epsilon_{\mathrm{th}}}^{+\infty}n(\epsilon)\,\mathrm{d}\epsilon\int_{-1}^{+1}\frac{1}{2}\,s\,(1-v_{p}\cos{\theta})\,\sigma_{p\gamma}(s)\,K(s)\,\mathrm{d}\cos{\theta}\,,
\end{align}
where $\sigma_{p\gamma}(s)$ is the proton-photon interaction cross section as a function of the squared center of mass energy (the Mandelstam $s$ variable), $n(\epsilon)$ represents the background photon density per unit volume and photon energy $\epsilon$, $\mu=\cos{\theta}$ is the impact parameter and $\epsilon_{\mathrm{th}}$ is the interaction threshold energy. $K(s)$ represents the reaction inelasticity, that is, the energy fraction available for secondary-particle production during the reaction. Complementary to the inelasticity is the elasticity function, defined as the energy fraction preserved by the primary particle after the interaction, $\eta={E_{\text{out}}}/{E_{\text{in}}}$ with the incoming particle energy $E_{\text{in}}$ and the residual energy $E_{\text{out}}$. Elasticity and inelasticity are connected by the simple relation: $K=(1-\eta)$.\\
The the Mandelstam $s$ can be computed introducing the photon four-momentum $(\epsilon',\,\vec{p}_{\gamma}')$ defined in the rest frame of the nucleus. In the high-energy limit approximation for the proton velocity $v_{p}\simeq1$, with $\mathrm{d}s=-2E_{p}\epsilon\,\mathrm{d}\cos{\theta}$, the following relations hold:
\begin{align}
\label{a43}
&s=(m_{p}+\epsilon')^2-|\vec{p}_{\gamma}'|^2=m_{p}^2+2m_{p}\epsilon'\\
&\epsilon'=\gamma\epsilon(1-v_{p}\cos{\theta})\,,
\end{align}
hence, the inverse of the attenuation length \cref{a42} can be written as:
\begin{equation}
\label{a44}
\frac{1}{l_{p\gamma}}=\frac{1}{2\,\gamma^{2}}\int_{\epsilon'_{\mathrm{th}}/2\gamma}^{+\infty}\mathrm{d}\epsilon \frac{n(\epsilon)}{\epsilon^2}\int_{\epsilon'_{\mathrm{th}}}^{\epsilon'_{\mathrm{max}}=2\gamma\epsilon}\frac{1}{2}\,\epsilon'\,\sigma_{p\gamma}(\epsilon')\,K(\epsilon')\,\mathrm{d}\epsilon'\,,
\end{equation}
where the primed quantities are defined in the proton rest frame, whereas the other quantities are defined in the laboratory rest frame. Since the $n(\epsilon)$ distribution is a Planckian function of the temperature $T$, a further simplification of the previous relation of \cref{a44} is possible, obtaining the explicit form for the inverse of the attenuation length \cite{Stecker}:
\begin{equation}
\label{a45}
\frac{1}{l_{p\gamma}}=-\frac{k_{\text{B}}\,T}{2\,\pi^2\,\gamma^2}\int_{\epsilon'_{\mathrm{th}}}^{+\infty}\epsilon'\,\sigma_{p\gamma}(\epsilon')\,K(\epsilon')\,\ln{\left(1-e^{-\epsilon'/2KT\gamma}\right)}\,\mathrm{d}\epsilon'\,.
\end{equation}
The inelasticity computed for the standard physics case is given by the relation \cite{Stecker1}:
\begin{equation}
\label{a46}
K(s)=\frac{1}{2}\left(1-\frac{m_{p}^{2}-m_{\pi}^{2}}{s}\right)\,.
\end{equation}

\section{QG introduced phenomenology in UHECR propagation}
In this work, following the theoretical framework introduced by HMSR, the kinematics of free particles is modified in order to include the QG effects. The kinematical perturbations modify the allowed phase space for the reaction and may therefore influence the processes involved in UHECR propagation. The introduction of QG phenomenology can indeed affect the photopion production mechanism, the core reaction underlying the GZK phenomenon \cite{Stecker,Scully,TorriUHECR,Torri:2020fao}. The QG-caused reduction of the allowed phase space can modify the inelasticity $K$ \cref{a46}. The consequent reduction of the inelasticity means that the incoming UHECR proton dissipates less energy during the GZK process and therefore the resulting opacity horizon is enlarged with respect to the standard physics prediction. In this work we conduct our analysis in the HMSR framework \cite{TorriHMSR}, but in the context of a version of the Finsler geometry obtained by starting from a MDR written in a more general form of \cref{a1} than the one given by \cref{a1b} and used in previous works \cite{TorriUHECR,Torri:2020fao}.

\subsection{Constraints from $\Delta$ resonance creation}
The photopion production requires a $\Delta$ particle creation (\cref{a41}) and can occur passing through a real $\Delta$ particle, in the case of the dominant process, or alternatively through a virtual one. The introduction of QG phenomenology can presumably modify only the GZK cut-off phenomenon, enlarging the foreseen opacity sphere without a complete suppression. The Universe opacity to the propagation of CR seems to be confirmed by experimental evidence, the uncertainty concerns only the dimension of the horizon. The production of the real $\Delta$ particle must therefore be preserved in order to foresee only small QG-induced deviations from the classical predictions for the GZK process. The four-momenta used in the following computations are considered as covariant vectors in order to simplify the use of the MLT for changing the reference frame. In the proton-CMB interaction process the production of a $\Delta$ particle is allowed if the proton-CMB interaction free energy, that is the Mandelstam $\sqrt{s}$ variable, exceeds the rest energy of the particle resonance. As a consequence of this threshold energy consideration, a first constraint on the magnitude of the QG perturbation parameters can be posed. Using the generalized internal product and the formalism developed for the modified kinematics in the \cref{a34,a35} one can write the Mandelstam $s$ variable as a function of the proton and the photon four-momenta respectively $p_{(p)}=(E_{(p)},\,\vec{p}_{(p)})$ and $p_{(\gamma)}=(E_{(\gamma)},\,\vec{p}_{(\gamma)})$:
\begin{align}
\label{a47}
s&=\left(p_{(p)}^{\mu}e_{\,\mu}^{a}(p_{p})+p_{\gamma}^{\mu}e_{\,\mu}^{a}(p_{\gamma})\right)\eta_{ab}\left(p_{(p)}^{\nu}e_{\,\nu}^{b}(p_{p})+p_{\gamma}^{\nu}e_{\,\nu}^{b}(p_{\gamma})\right) \notag \\
&=\;p_{(p)}^{\mu}e_{\,\nu}^{a}(p_{p})\,\eta_{ab}\,e_{\,\nu}^{b}(p_{p})p_{(p)}^{\nu}+p_{(p)}^{\mu}e_{\,\nu}^{a}(p_{p})\,\eta_{ab}\,e_{\,\nu}^{b}(p_{\gamma})p_{(\gamma)}^{\nu} \notag \\
&\phantom{{}={}}+p_{(\gamma)}^{\mu}e_{\,\nu}^{a}(p_{\gamma})\,\eta_{ab}\,e_{\,\nu}^{b}(p_{p})p_{(p)}^{\nu} \notag \\
&=\;p_{(p)}^{\mu}g_{\mu\nu}(p_{p})p_{(p)}^{\nu}+2p_{(p)}^{\mu}e_{\,\mu}^{a}(p_{p})\,\eta_{ab}\,\delta_{\,\nu}^{b}\,p_{(\gamma)}^{\nu}\geq m_{\Delta}^{2}\,.
\end{align}
The vierbein \cref{a14} has been used to project the momenta of different particles to the common support Minkowski spacetime and the symmetry of the mixed product of $p_{p}$ and $p_{\gamma}$ has been used. Photons are assumed to be Lorentz-invariant, hence the related vierbein is defined as usual: $e_{\mu}^{a}(p_{\gamma})=\delta_{\mu}^{a}$. Using the explicit form of the squared Finsler norm $F^2(p)$ given in \cref{a15ax}, the following inequality can be derived reordering the terms of the previous relation in an opportune manner:
\begin{align}
\label{a47a}
&E_{p}^{2}(1-h_{p}(p_{p}))-|\vec{p}_{p}|^{2}(1-k_{p}(p_{p}))+2E_{p}E_{\gamma}\left(1-\frac{1}{2}h_{p}(p_{p})\right) \notag \\
&-2\vec{p}_{p}\cdot\vec{p}_{\gamma}\left(1-\frac{1}{2}k_{p}(p_{p})\right)\geq m_{\Delta}^{2}\,,
\end{align}
where $h_{p}(p_{p})$ and $k_{p}(p_{p})$ are the proton QG perturbation functions taken from the MDR \cref{a14}. The following inequality can be derived from the previous \cref{a47a} using the MDR \cref{a1}:
\begin{equation}
\label{a48}
2\left(h_{p}(p_{p})-k_{p}(p_{p})\right)E_{p}^{2}-E_{p}E_{\gamma}\bigg(4+\underbrace{(h_{p}(p_{p})-k_{p}(p_{p}))}_{O(1)}\bigg)+m_{\Delta}^{2}-m_{p}^{2}\leq0\,,
\end{equation}
where $m_{\Delta}\simeq \unit[1232]{MeV}$, $m_{p}\simeq \unit[938]{MeV}$. The covariant formulation of the model is fundamental since it allows the change of the reference frame. Indeed, in order to simplify the computation, it is possible to choose the more suitable reference frame, where the involved energy scales ratio allows the suppression of some perturbative terms. One can therefore consider the laboratory frame, where the proton energy is much higher than the CMB one. Considering that the UHECR proton energy has an upper limit $E_{p}\lesssim \unit[10^{21}]{eV}$ and taking into account the tiny average value of the CMB energy $E_{\gamma}\simeq \unit[1.16\times10^{-3}]{eV}$, one can neglect the $O(1)$ contributions in the second term obtaining:
\begin{equation}
\label{a48a}
2\left(h_{p}(p_{p})-k_{p}(p_{p})\right)E_{p}^{2}-4E_{p}E_{\gamma}+m_{\Delta}^{2}-m_{p}^{2}\leq0\,.
\end{equation}
Defining the function $f_{p}=h_{p}-k_{p}$ the result obtained is comparable with the one presented in \cite{TorriUHECR,Torri:2020fao}. The derived inequality must be satisfied in order to produce a $\Delta$ resonance, otherwise the QG perturbations totally suppress the GZK effect. Imposing the validity of the previous relation the following constraint can be derived:
\begin{equation}
\label{a49}
f_{p}(p_{p})\leq\frac{2E_{\gamma}}{m_{\Delta}^2-m_{p}^2}\,.
\end{equation}
Substituting the approximated average value of the CMB energy in the previous inequality one can obtain the approximated constraint:
\begin{equation}
\label{a50}
f_{p}(p_{p})\lesssim 10^{-23}\,.
\end{equation}
This result is comparable with the upper limit $4.5\cdot10^{-23}$ obtained numerically in \cite{Stecker} for the perturbation magnitude.\\
As a final remark from the relations \cref{a48,a49} it follows that in order to generate visible effects on the GZK phenomenon, the QG perturbation function $f_{p}$ must satisfy the relation:
\begin{equation}
\label{a51}
f_{p}(p_{p})=h_{p}(p_{p})-k_{p}(p_{p})>0\,.
\end{equation}
Indeed from \cref{a48} the high-energy limit excludes the possibility of any suppression of the GZK cut-off in the case of a negative function $f_{p}$. We underline that the choice of positive values of the perturbation $f_{p}$ corresponds to the introduction of an effective MAV in the MDR of \cref{a1} for every massive particle, which is lower than the standard speed of light.

\subsection{Modified inelasticity}
The introduction of QG perturbations can modify the phase space allowed for the photopion production process, determining a modification of the inelasticity function $K(s)$ \cref{a46}.  In this work we evaluate the new inelasticity in the context of the generalized Finsler geometry here introduced following the approach of \cite{TorriUHECR,Torri:2020fao}, originally inspired by \cite{Stecker,Scully}.
In the following, all the computations are conducted again considering the four-momenta as covariant vectors, in order to simplify the computations transforming the reference frame with the MLT. The introduction of the center-of-mass (CM) reference frame is useful and it is defined via the relation:
\begin{equation}
\label{a52}
\vec{p}_{p}^{*}+\vec{p}_{\pi}^{*}=0\,,
\end{equation}
where the momenta are defined on the common Minkowski support spacetime $(TM,\,\eta_{ab})$ and $*$ labels the quantities related to the CM frame.\\
The next element necessary in evaluating the change from the CM to a generic reference frame is the $\gamma_{\mathrm{CM}}$ Lorentz factor. Starting from the free energy available for the photopion production $\sqrt{s}=E_{p}^{*}+E_{\pi}^{*}$ one can obtain the relation:
\begin{equation}
\label{a53}
\gamma_{\mathrm{CM}}(E_{p}^{*}+E_{\pi}^{*})=\gamma_{\mathrm{CM}}\sqrt{s}=(E_{p}+E_{\pi})\;\Rightarrow\;\gamma_{\mathrm{CM}}=\frac{E_{p}+E_{\pi}}{\sqrt{s}}=\frac{E_{\mathrm{tot}}}{\sqrt{s}}\,.
\end{equation}
Using the CM definition \cref{a52} the relation $|\vec{p}^{*}_{p}|=\left|\vec{p}^{*}_{\pi}\right|$ follows and the four-momentum of the photopion can be written in the CM reference frame as $p_{\pi}^{*}=(\sqrt{s}-E^{*}_{p},\,\vec{p}_{p}^{*})=(E^{*}_{\pi},\,\vec{p}_{\pi}^{*})$. From the latter, the free energy necessary to produce a photopion in the CM frame can be computed using the squared Finsler norm \cref{a15ax}:
\begin{align}
\label{a54}
&p_{(\pi)}^{\mu}e_{\,\mu}^{a}(p_{\pi})\,\eta_{ab}\,e_{\,\nu}^{b}(p_{\pi})p_{(\pi)}^{\nu}=m_{\pi}^{2} \notag \\
\Rightarrow &\;\big(\sqrt{s}-E^{*}_{(p)}\big)^{2}\big(1-h_{\pi}(p_{\pi})\big)-|\vec{p}^{*}_{(p)}|\big(1-k_{\pi}(p_{\pi})\big) \notag \\
&=\;\big(s-2\sqrt{s}E^{*}_{(p)}\big)\big(1-h_{\pi}(p_{\pi})\big)+E_{(p)}^{*2}\big(1-h_{p}(p_{p})\big)+E_{(p)}^{*2}\big(h_{p}(p_{p})-h_{\pi}(p_{\pi})\big) \notag \\
&\phantom{{}={}}-\;|\vec{p}^{*}_{(p)}|^{2}\big(1-k_{p}(p_{p})\big)-|\vec{p}^{*}_{(p)}|^{2}\big(k_{p}(p_{p})-k_{\pi}(p_{\pi})\big) \notag \\
&=\;\big(s-2\sqrt{s}E^{*}_{(p)}\big)\big(1-h_{\pi}(p_{\pi})\big)+m_{p}^{2}\big(1-h_{\pi}(p_{\pi})\big)+E_{(p)}^{*2}\big(h_{p}(p_{p})-h_{\pi}(p_{\pi})\big) \notag \\
&\phantom{{}={}}-\;|\vec{p}^{*}_{(p)}|^{2}\big(k_{p}(p_{p})-k_{\pi}(p_{\pi})\big)=m_{\pi}^{2}\,,
\end{align}
$h_{p}$ and $k_{p}$ are the perturbation functions of the proton MDR of \cref{a1} and $h_{\pi}$ and $k_{\pi}$ are the analogous functions for the pion MDR, $p$ and $\pi$ label, respectively, the elements related to the proton and the pion.\\
We underline that the computation made here for the modifications induced on the inelasticity are still valid in the context of the more restrictive Finsler geometry. Indeed, the computation is conducted using the MDR \cref{a1}, that is the norm \cref{a15ax}.\\
From the previous equation and neglecting the $O(2)$ perturbations in \cref{a15ax}, the following equality can be obtained:
\begin{align}
\label{a55}
E_{p}^{*}&=\frac{s+(m_{p}^2-m_{\pi}^2)(1+h_{\pi})+E_{p}^{*2}(h_{p}-h_{\pi})-|\vec{p}^{*}_{p}|^{2}(k_{p}-k_{\pi})}{2\sqrt{s}} \notag \\
&=\;\frac{s+(m_{p}^2-m_{\pi}^2)(1+h_{\pi})+h_{p\pi}E_{p}^{*2}-k_{p\pi}|\vec{p}^{*}_{p}|^{2}}{2\sqrt{s}}=F(s)\,,
\end{align}
where $h_{p\pi}=h_{p}-h_{\pi}$ and $k_{p\pi}=k_{p}-k_{\pi}$ are the QG perturbation parameters. Considering the high-energy limit $E_{p}^{*}\simeq|\vec{p}^{*}_{p}|$ in the previous relation the residual proton energy after the photopion reaction can be increased if the QG parameters satisfy the inequality $h_{p\pi}>k_{p\pi}$ and therefore the relation $h_{p\pi}>0$. The proton-dominant correction $h_{p}$ must be larger than the pion one $k_{p}$ in order to produce a dilatation of the GZK sphere. In this work we assume the gravitational nature of the perturbation effects caused by the supposed quantum structure of the background, hence the heavier particles have a bigger QG-induced modification. As a consequence, the correction factors of the pion can be considered negligible and it is possible to pose $h_{p\pi}\simeq h_{p}$ and $k_{p\pi}\simeq k_{p}$.\\
Now the following approximations can be introduced:
\begin{subequations}
\label{a56}
\begin{align}
p^{*}_{p}\simeq E^{*}_{p}&=(1-K_{\pi}(\theta))\sqrt{s} \\[1ex]
p^{*}_{\pi}\simeq E^{*}_{\pi}&=K_{\pi}(\theta)\sqrt{s}\,.
\end{align}
\end{subequations}
$E=\sqrt{s}$ represents the initial free total energy and $E'_{p}$ and $E'_{\pi}$ are the final energies of the proton and the pion, respectively.\\
Using the Lorentz transformations for changing reference frames, one can obtain the following:
\begin{subequations}
\label{a57}
\begin{align}
E'_{p}&=\gamma_{\mathrm{CM}}(E_{p}^{*}+\beta \cos{\theta} p^{*}_{p})\\[1ex]
E^{*}_{p}&=(1-k_{\pi}(\theta))\sqrt{s}\,,
\end{align}
\end{subequations}
where $K_{\pi}(\theta)$ is the pion inelasticity depending on the impact angle.\\
In the high-energy limit, where for ultra-relativistic particles $|\vec{p}|\sim E$ and the velocity factor $\beta\sim1$ and $\gamma_{\mathrm{CM}}$ is approximated by \cref{a53}, the following equation can be derived:
\begin{equation}
\label{a58}
(1-K_{\pi}(\theta))=\frac{1}{\sqrt{s}}\left(F(s)+\cos{\theta}\sqrt{F(s)^2-m_{p}^2+2(h_{p}-k_{p})|\vec{p}|^2}\right)\,.
\end{equation}
Posing $f_{p}=(h_{p}-k_{p})$ in the previous relation, it is possible to obtain the following one:
\begin{equation}
\label{a59}
(1-K_{\pi}(\theta))=\frac{1}{\sqrt{s}}\left(F(s)+\cos{\theta}\sqrt{F(s)^2-m_{p}^2+2f_{p}|\vec{p}|^2}\right)
\end{equation}
where the latter result was obtained in \cite{TorriUHECR,Torri:2020fao}. The previous equation can be numerically solved in order to evaluate the inelasticity as a function of the collision angle $\theta$. The numerical computation is conducted in the high-energy limit, hence by using \cref{a56} and considering $\sqrt{s}=E$ where $E$ is the initial energy for the process.\\
The free energy $s$ can be written as:
\begin{equation}
\label{a60}
s=(E_{p}+\epsilon')^2-\left|\vec{p}_{\gamma}\right|^2=2m_{p}\epsilon'+m_{p}^2\,,
\end{equation}
introducing the energy of the photon $\epsilon'$ defined in the proton rest frame $\vec{p}_{p}=0$.\\
The inelasticity is finally averaged over the interval $\theta\in[0,\;\pi]$:
\begin{equation}
\label{a61}
K_{\pi}=\frac{1}{\pi}\int_{0}^{\pi}K_{\pi}(\theta)\;\mathrm{d}\theta\,.
\end{equation}
In the following, we plot the inelasticity for different choices of the parameter $f_{p}=(h_{p}-k_{p})$. The parameter $f_{p}$ is constrained to be positive in order to guarantee detectable QG effects. This hypothesis corresponds to the introduction of a MAV inferior to the classical speed of light $c$ for every massive particle species. The parameter is even constrained from above by the limit obtained in the previous section \cref{a50}, hence the plausible parameter must be included in the interval $0<f_{p}\lesssim10^{-22}$. The inelasticity is plotted as a function of the proton energy $E_{p}$, defined in the laboratory reference frame, and the photon energy $\epsilon'$, defined in the proton rest frame. Under these hypotheses, the QG perturbations cause a dramatic drop of the inelasticity value, which is more visible for increasing values of the parameter $f_{p}$. This implies a reduction of the allowed phase space for the photopion production process, and an enlargement of the foreseen opacity horizon for an increasing magnitude of the QG perturbation.

\section{Simulated attenuation lenght}
\begin{figure}[H]
\includegraphics[scale=0.56]{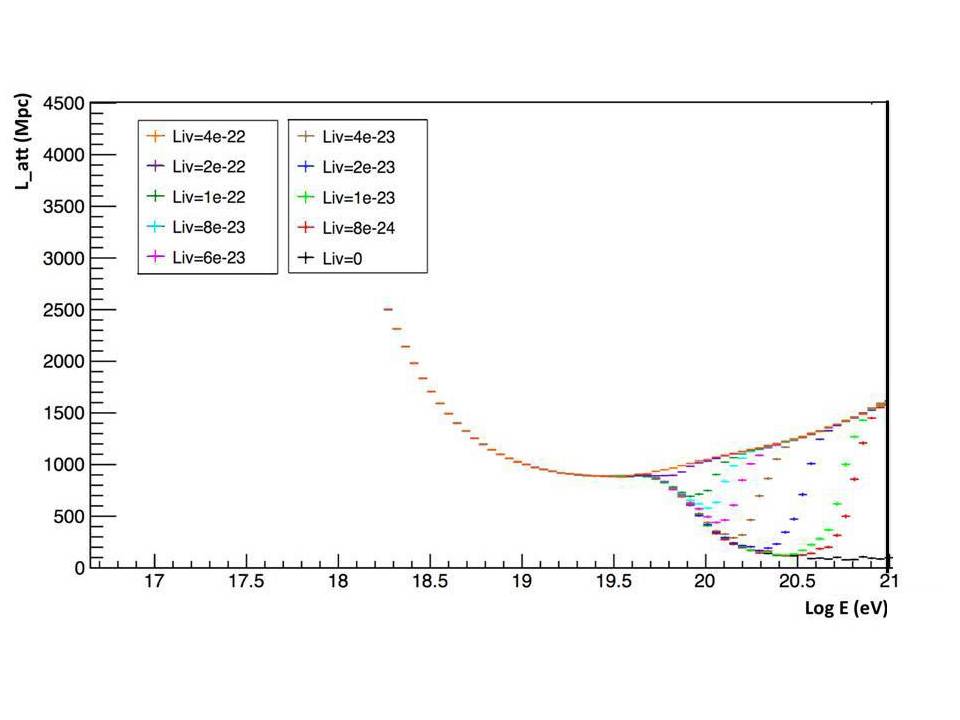}
\caption{Attenuation length as a function of energy, plotted for ten different values of the LIV parameter.}
\label{fig:Latt1}
\end{figure}

\begin{figure}[H]
\includegraphics[scale=0.49]{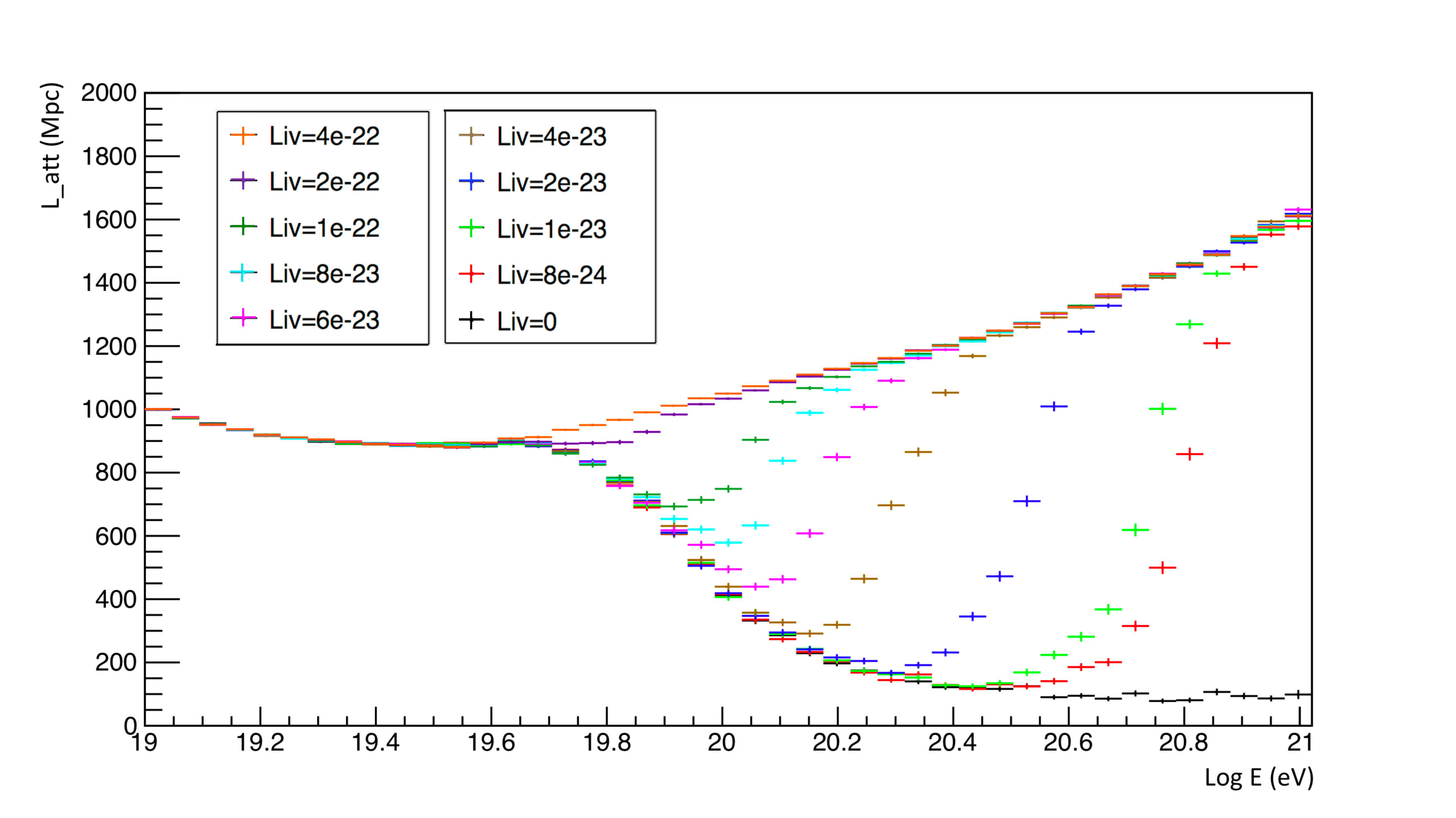}
\caption{Attenuation length as a function of energy; blow up of the previous plot.}
\label{fig:Latt2}
\end{figure}

The impact of the QG perturbations on the UHECR free propagation can be evaluated computing the value of the attenuation length as a function of the modified inelasticity $K$ in \cref{a45}. In this work we have obtained the attenuation length using an ad hoc modified version of the simulation software \emph{SimProp} \cite{Aloisio}. In this software version we substituted the inelasticity predicted by the standard physics with the modified one, being a function not only of the proton and the CMB energy but also of the QG parameter. In Figs.~\ref{fig:Latt1} and \ref{fig:Latt2} it is visible that the increase of the opacity horizon can be caused by the QG-induced modification of Lorentz covariance. In the QG-less scenario the attenuation length decreases for increasing CR energy values. In the presence of QG, the interaction length at first decreases with the energy, but then after an inflection point, which depends on the energy, this quantity starts rising for increasing energy values. This effect is caused by the reduction of the average energy lost for every proton-CMB interaction caused by the reduction of the inelasticity. For a QG parameter of $f_{p}\sim10^{-22}$ the modifications in the attenuation length are appreciable starting from an energy $E\sim \unit[10^{20}]{eV}$ of the incoming proton, instead for a parameter $f_{p}\sim2\times10^{-22}$ the perturbation starts being appreciable yet at an energy $E\sim \unit[6\times10^{19}]{eV}$. The analysis can be improved taking into account even the electron-positron pair production. This effect is not dominant at the highest energies, but would presumably further increase the energy loss process.

\begin{figure}[H]
\begin{center}
\includegraphics[scale=0.35]{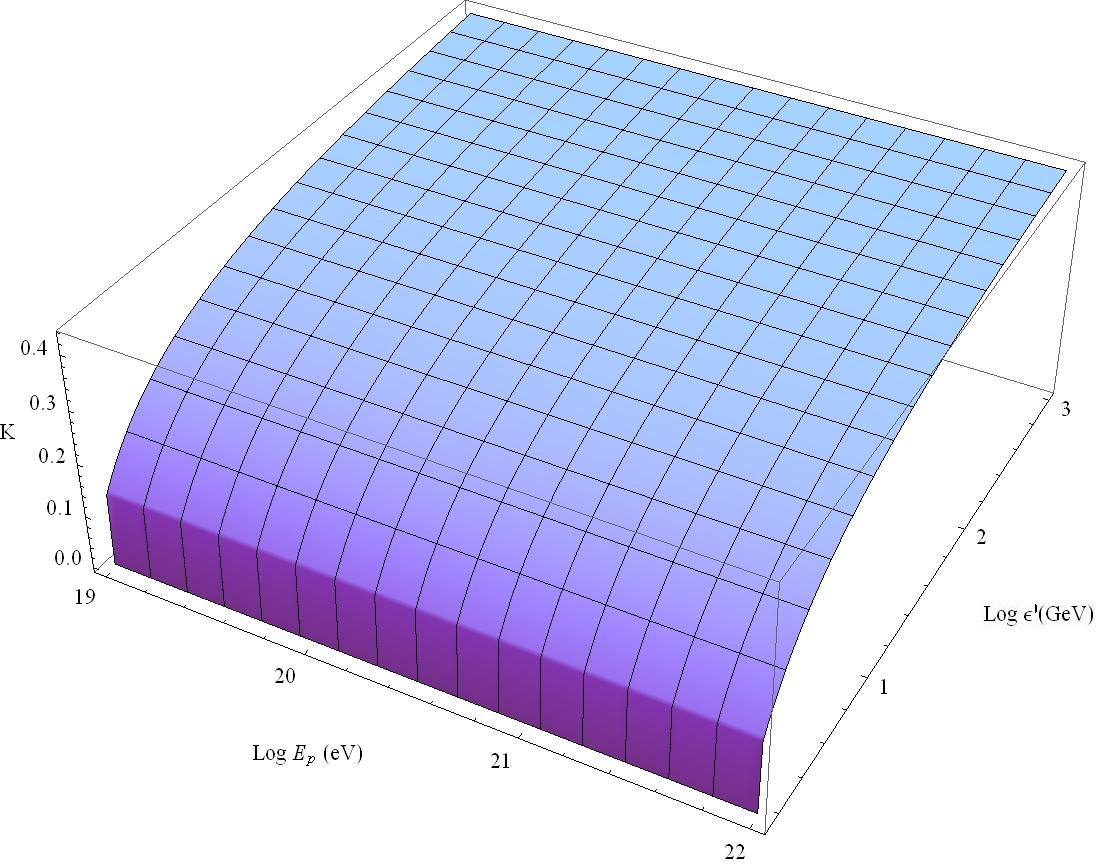}
\includegraphics[scale=0.30]{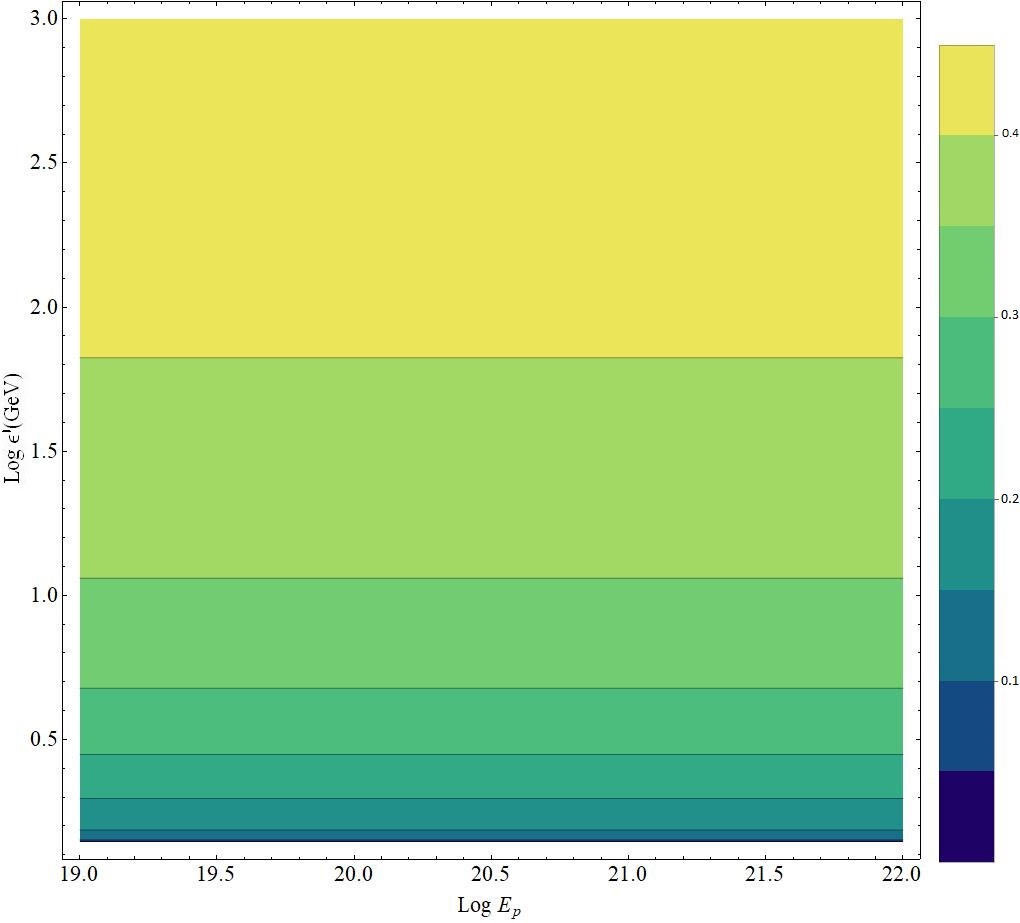}
\caption{Inelasticity obtained in the case of the LIV parameter $f_{p\pi}\simeq f_{p}=0$ as a function of the proton energy $E_{p}$ and of the photon energy $\epsilon'$ defined in the proton rest frame.}
\label{fig:inel1}
\end{center}
\end{figure}

\begin{figure}[H]
\begin{center}
\includegraphics[scale=0.35]{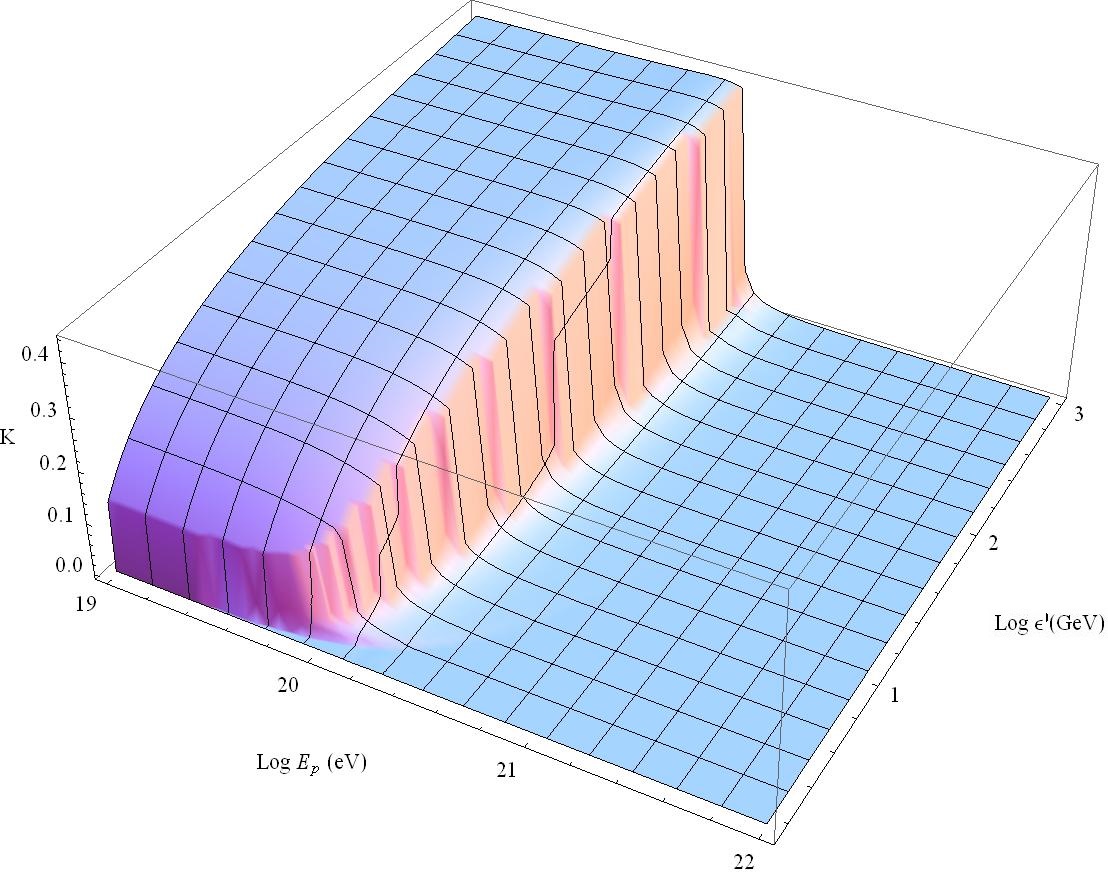}
\includegraphics[scale=0.30]{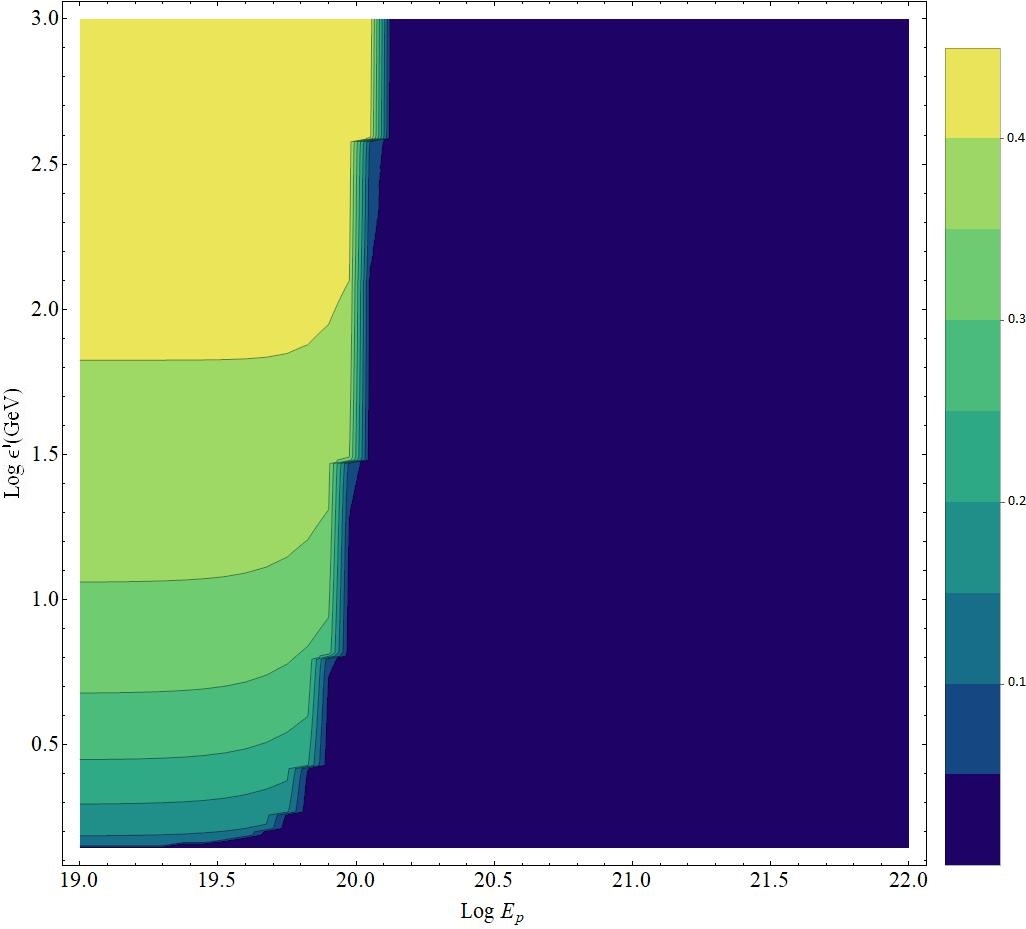}
\caption{Inelasticity obtained in the case of the LIV parameter $f_{p\pi}\simeq f_{p}=9\times10^{-23}$ as a function of the proton energy $E_{p}$ and of the photon energy $\epsilon'$ defined in the proton rest frame.}
\label{fig:inel2}
\end{center}
\end{figure}

\begin{figure}[H]
\begin{center}
\includegraphics[scale=0.35]{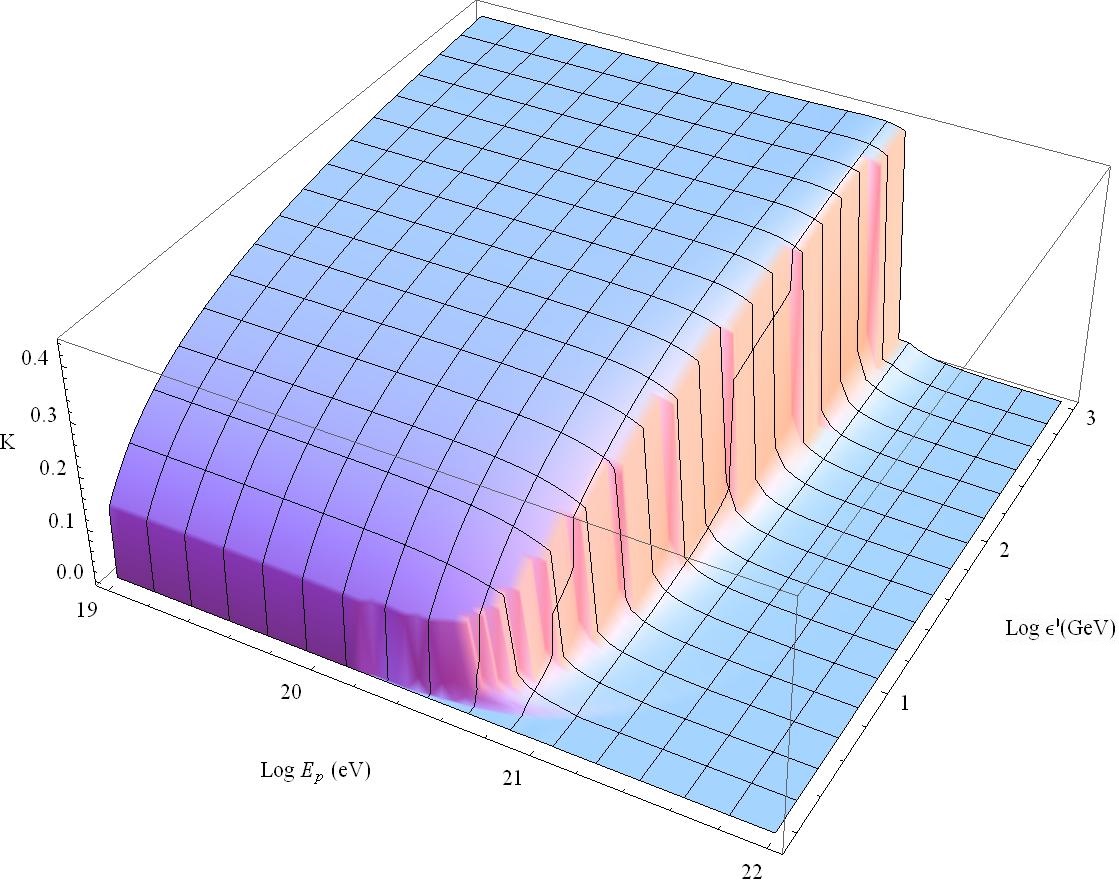}
\includegraphics[scale=0.30]{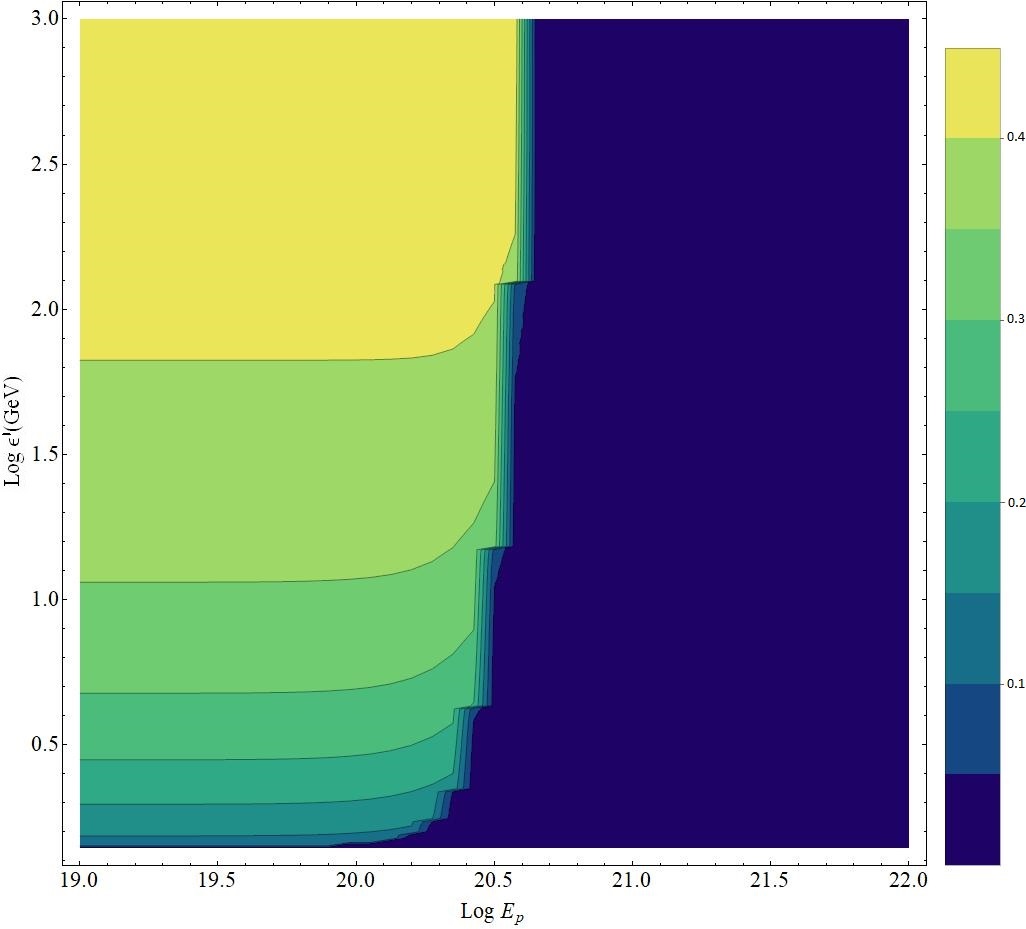}
\caption{Inelasticity obtained in the case of the LIV parameter $f_{p\pi}\simeq f_{p}=3\times10^{-24}$ as a function of the proton energy $E_{p}$ and of the photon energy $\epsilon'$ defined in the proton rest frame.}
\label{fig:inel3}
\end{center}
\end{figure}

\section{Conclusions}
In this work we investigated the possibility of exploiting the UHECR physics to detect supposed signatures of the quantum structure of spacetime. Nowadays the Lorentz covariance stands at the base of our physics knowledge, but small departures can be induced by QG effects. We studied the induced QG phenomenology on UHECRs simulating their propagation in a framework of Lorentz covariance modification. The model is a generalization of the HMSR theory, that foresees a minimal extension of the particle SM, preserves the usual gauge symmetry $SU(3)\otimes SU(2)\otimes U(1)$, and does not introduce exotic particles or reactions. The spacetime geometry is a generalized Finsler spacetime and all the GZK cutoff modification effects have been computed in this context. We underline that the methodology introduced in this work can be used to set the stage of the model in the context of the more restrictive Finsler geometry. Moreover, we have demonstrated that, for instance, the inelasticity modification that underlies the opacity sphere enlargement can be still computed in the Finsler geometric spacetime.\\
The conducted simulations foresee an enlargement of the attenuation length in accordance with other works \cite{Stecker,Scully,TorriUHECR,Torri:2020fao,Bietenholz:2008ni,Mattingly:2005re}, but are obtained in a covariance-preserving scenario as proposed in \cite{AmelinoCamelia:2002gv}. The kinematical solution proposed in our work is based on the phenomenological effects caused by the interaction of different particle species which are modified in different ways by their interactions with the QG structure of spacetime. Our proposal is analogous to that of \cite{Coleman}, but is based on a covariant formulation, such as in DSR theories \cite{Amelino-Camelia:2019cjb,AmelinoCamelia:2011er}. Moreover, in our proposal a way is outlined to generalize HMSR theory in curved spacetime and a formalism is introduced that defines the threshold energy of reactions in a modified covariant framework.\\
We emphasize the importance for this work of preserving covariance, even if in an amended formulation. The preserved covariance allows to define relativistic kinematical invariants. As a direct consequence, it is possible to choose the most suitable reference frame to simplify the computations, suppressing some perturbation terms thanks to the involved energy scale ratios. Furthermore, in the astroparticle sector covariance can be a great experimental advantage, since all the obtained data can be collected without the necessity to introduce any sidereal discrimination related to the orientation of the detector with respect to a fixed privileged reference frame.\\
In this work we set the stage for an analysis strategy that can be improved including a heavier UHECR component, which is a more realistic CR composition, as indicated by the Pierre Auger collaboration \cite{Aab:2014aea}. The QG perturbations are expected to be larger for a CR component heavier than protons. Indeed, the kinematical perturbations can affect the propagation of bare nuclei in a more significant way.


\vspace{18pt}




\textbf{Funding}
This work was supported by the Fondazione Fratelli Giuseppe Vitaliano, Tullio e Mario
Confalonieri—Milano, financing a post doctoral fellowship at Dipartimento di Fisica - Università degli Studi di Milano.
\bigskip

\textbf{Acknowledgments}\\
The author wants to thanlk Christian Pfeifer and Nicoleta Voicu for the useful discussion and the advices to improve this work.

\textbf{Conflicts of interest}\\
The authors declare no conflict of interest.
\bigskip


\textbf{Abbreviations}
The following abbreviations are used in this manuscript:
\begin{itemize}
  \item \textbf{UHECR} Ultra High Energy Cosmic Ray
  \item \textbf{CR} Cosmic Rays
  \item \textbf{SR} Special Relativity
  \item \textbf{GR} General Relativity
  \item \textbf{QG} Quantum Gravity
  \item \textbf{LIV} Lorentz Invariance Violation
  \item \textbf{HMSR} Homogeneously Modified Special Relativity
  \item \textbf{DSR} Doubly Special Relativity
  \item \textbf{SM} Standard Model
  \item \textbf{SME} Standard Model extension
  \item \textbf{VSR} Very Special Relativity
  \item \textbf{DR} Dispersion Relation
  \item \textbf{QFT} Quantum Field Theory
  \item \textbf{MDR} Modified Dispersion Relation
  \item \textbf{MLT} Modified Lorentz Transformation
\end{itemize}

\bigskip

\appendix
\section{}
In this appendix we show that it is possible to choose perturbations $h$ and $k$, functions of the ratio $|\vec{p}|/E$, whose derivatives exhibit again a perturbative character.
If we admit for these functions the general form:
\begin{equation}
h\left(\frac{|\vec{p}|}{E}\right)=\sum_{n}\alpha_{n}\left(\frac{|\vec{p}|}{E}\right)^{n} \qquad k\left(\frac{|\vec{p}|}{E}\right)=\sum_{n}\beta_{n}\left(\frac{|\vec{p}|}{E}\right)^{n}\,,
\end{equation}
where the function magnitude is encoded in the coefficients $\alpha_{n}\ll1$ and $\beta_{n}\ll1$, hence it is possible to choose these coefficients in order to pose: $h$, $k\simeq O(1)$.\\
The definition domain of $h$ and $k$ is the interval $[0,\,1)$. Hence, requiring a strongly limited magnitude, even in the high-energy limit $|\vec{p}|,\,E\rightarrow+\infty$, one obtains the convergence of the series:
\begin{equation}
\sum_{n}\alpha_{n}<+\infty, \qquad \sum_{n}\beta_{n}<+\infty.
\end{equation}
Therefore, the sequences $\alpha_{n},\,\beta_{n}$ must decrease fast enough admitting the limit
\begin{equation}
\alpha_{n},\,\beta_{n}\rightarrow0.
\end{equation}
The derivatives of the functions $h$ and $k$ with respect to the variables $p_{i}$ and $E$ take the following form:
\begin{subequations}
\begin{align}
\frac{\partial h(p)}{\partial p_{i}}&=\sum_{n}n\,\alpha_{n}\frac{p_{i}|\vec{p}|^{n-1}}{E^{n}} \\[1ex]
\frac{\partial h(p)}{\partial E}&=-\sum_{n}n\,\alpha_{n}\frac{|\vec{p}|^{n}}{E^{n+1}}\,,
\end{align}
\end{subequations}
where $i\in\{1,\,2,\,3\}$.\\
A simple check of the assertion can be obtained considering the $h$ and $k$ as linear, that is, series ending at the first expansion term. In this case the magnitude order of the functions and their respective derivatives is determined by the coefficients $\alpha_{1}$ and $\beta_{1}$ and therefore present the same perturbation character.\\
In the more general case the assertion can be proved as a direct consequence of these last equations, indeed it is possible to choose the coefficients $\alpha_{n}$ and $\beta_{n}$ such that even the derivatives of the functions $h$ and $k$ are limited in the definition domain and their magnitudes are determined again by the first coefficients of the expansions, since they belong to quickly decreasing sequences. The chosen perturbation functions $h$ and $k$ admit, therefore, derivatives with the same magnitude, proving the assertion. Some explicit examples are the linear function $\alpha(|\vec{p}|/E)$, the truncated exponential series $\exp({\alpha|\vec{p}|/E})-1$ and the trigonometric function $\sin({\alpha|\vec{p}|/E})$, with an ad hoc choice for the coefficient $\alpha$, which must be strongly limited and determines the perturbative character of the selected function.\\
The perturbative order of the matrix $A$ follows from a direct computation. Indeed, the entries of the matrix $A$ involve first and second-order derivatives of the perturbation functions $h$ and $k$ multiplied, respectively, by the momentum to the first or the second power.

\bigskip
\bigskip
\end{document}